\documentclass[prd,groupaddress,nofootinbib]{revtex4}

\usepackage{slashed}
\usepackage{graphicx}
\usepackage{amsmath}
\usepackage{amssymb}
\usepackage{epsfig}
\usepackage{hhline}
\usepackage{soul}
\usepackage{tabularx,pifont,multirow}
\usepackage{enumitem}
\usepackage{colordvi}
\usepackage{soul}
\usepackage{xcolor}
\usepackage{yfonts}

\definecolor{darkgreen}{rgb}{0,0.3,0.05}
\makeatletter                                                         %
\newcommand*\rel@kern[1]{\kern#1\dimexpr\macc@kerna}                  %
\newcommand*\widebar[1]{                                              %
  \begingroup                                                         %
  \def\mathaccent##1##2{                                              %
    \rel@kern{0.8}                                                    %
    \overline{\rel@kern{-0.8}\macc@nucleus\rel@kern{0.2}}             %
    \rel@kern{-0.2}                                                   %
  }                                                                   %
  \macc@depth\@ne                                                     %
  \let\math@bgroup\@empty \let\math@egroup\macc@set@skewchar          %
  \mathsurround\z@ \frozen@everymath{\mathgroup\macc@group\relax}     %
  \macc@set@skewchar\relax                                            %
  \let\mathaccentV\macc@nested@a                                      %
  \macc@nested@a\relax111{#1}                                         %
  \endgroup                                                           %
}                                                                     %
\makeatother                                                          %

\include{Kaons_macro}

\begin{document}
\preprint[\leftline{KCL-PH-TH/2014-{\bf 30}, \, LCTS/2014-43}

%

\title{\Large {\bf Tree level Leptogenesis from Kalb-Ramond Torsion Background } }

\bigskip

\author{M. de Cesare}

\author{Nick E.~Mavromatos}\altaffiliation{Currently also at: Theory Division, Physics Department, CERN, Geneva 23,
  CH 1211, Switzerland}
  
\author{Sarben Sarkar}

\affiliation{\vspace{1mm}
Theoretical Particle Physics and Cosmology Group,
  Department of Physics, King's College London, Strand, London WC2R
  2LS, UK}


\begin{abstract}
\vspace{0.5cm}
\centerline{\bf Abstract }
\noindent\\[-2mm] 
The  effect of  torsion  in theories  of  quantum gravity  is
known to be well described by  an axion-like field which couples  to matter as well as 
to gravitation  and radiation  gauge fields.  In this note we consider a particular kind of torsion, arising from the Kalb-Ramond antisymmetric 
tensor field that appears in the gravitational multiplet of string theory. We investigate the implications for leptogenesis. It is shown that leptogenesis can occur even at tree-level and with only one generation of right-handed Majorana neutrinos, due to CP and CPT violation introduced by the background geometry.
\end{abstract}
\maketitle

%
%
%
%
%
\newcommand{\be}{\begin{equation}}
\newcommand{\ee}{\end{equation}}
\newcommand{\eps}{\varepsilon}
\newcommand{\slas}{\slashed}

\section{Introduction}

\label{s1}
Baryogenesis represents a long-standing problem and is very much an active research area in modern cosmology.
In fact any theory of cosmology that does not explain baryogenesis can be considered as incomplete.The current proposals for explaining baryogenesis highlight the need for additional mechanisms for the generation of a baryon asymmetry. We will use the gravitational sector of string theory to give a new mechanism for baryogenesis via leptogenesis.

A solution for baryogenesis would explain why the primordial Universe, that was dominated by radiation, evolved into the present matter dominated Universe. Many approaches, proposed in the literature, are reviewed in  \cite{{Mohapatra:1998rq,Trodden,Buchmuller review,Chen:2007fv,pilaftsis}}.  
A standard measure of the abundance of baryons over that of antibaryons is  defined by the ratio \cite{Buchmuller-Yanagida}
\be \label{baryon asymmetry}
Y_{\Delta B}=\frac{n_{B}-n_{\bar{B}}}{n_{\gamma}}=(6.1\pm0.3)\times10^{-10}
 \ee
where $n_{B}$ is the number density of baryons, $n_{\bar{B}}$ is
the number density of antibaryons and $n_{\gamma}$ is the density of photons (proportional to the entropy density $s$).
This number was determined with accurate measurements of the CMB radiation by the experiments WMAP \cite{WMAP} and Planck \cite{Planck}.
However, there is no experimental evidence for primordial antimatter in the visible universe.
 Similarly, the generation of an asymmetry between leptons.
and antileptons is known as leptogenesis. This is expected to be of the same order of magnitude as $Y_{\Delta B}$.

If $B$, the net baryon number,  is conserved in Nature, the matter asymmetry can only originate from an asymmetric initial condition $B\neq0$. However, such an asymmetry would rapidly  diminish during inflation, and extreme fine tuning of the initial condition would become necessary. This is highly unsatisfactory from a theoretical point of view. Consequently a mechanism for the dynamical generation of a baryon asymmetry is required.

In the seminal paper \cite{Sakharov}, Sakharov first identified a set of three necessary conditions that must be satisfied in order to produce a net baryon number. 
\begin{enumerate}
\item The theory must allow for interactions that violate $B$ conservation. These interactions must become effective at high energy scales in order to guarantee the stability of the proton.
\item Both discrete symmetries $C$ (charge conjugation) and $CP$ (where $P$ denotes parity) are violated. In fact $C$ violation is not enough, as correlations between the spins of particles and antiparticles lead to identical cross sections for conjugated processes~\cite{Zeldovich} when  the theory is $CP$ symmetric.
\item  A departure from thermal equilibrium must occur: a $CPT$ invariant theory (where $T$ denotes time reversal) does not allow $\langle B\rangle\neq0$ at thermal equilibrium.  
\end{enumerate}
A detailed review of Sakharov's conditions in different baryogenesis models can be found in \cite{Trodden, Buchmuller review}.
The third Sakharov condition implicitly assumes that the underlying field theory is  invariant under the discrete symmetry operator $\Theta\equiv CPT$. This assumption is usually valid due to the CPT theorem \cite{Wightman}: $\Theta$ is an invariance of local Lorentz invariant quantum field theories. $\Theta$ invariance is not always valid, for example (i) in models of spontaneous baryogenesis (see \emph{e.g.} \cite{Cohen Kaplan}, \cite{Cohen Kaplan 2}) and (ii)  through interactions with external fields \cite{Bertolami-Colladay-et al.} where the matter asymmetry is produced in equilibrium. Recently it was emphasised by Greenberg that CPT violation also implies Lorentz violation \cite{Greenberg}.

On closer inspection the Standard Model (SM) can be seen to satisfy the Sakharov conditions: 
\begin{itemize}
\item At the classical level the Lagrangian of SM has global
$U\left(1\right)$ chiral symmetries, which lead to $B$ conservation as
well as $L$ conservation for individual generations. The gauge group
of SM is the direct product group $SU(3)_{c}\times SU\left(2\right)_{w}\times U(1)_{Y}$.
At the quantum level, however, the currents of these global
symmetries are anomalous \cite{'t Hooft}.  This anomaly, known as the triangle anomaly \cite{anomalies, Adler, Bell-Jackiw}, appears since left chiral and right chiral fields interact
differently with the gauge fields $W_{\mu}^{a}$ (associated with
$SU\left(2\right)_{w}$ ) and $B_{\mu}$ (associated with $U(1)_{Y}$).            
The currents associated with $B$ and $L$ are denoted by $j_{B}^{\mu}$ and
$j_{L}^{\mu}$ respectively; for the case of $N_{f}$ fermion generations,
the triangle anomaly implies that 
\begin{equation}
\partial_{\mu}J_{B}^{\mu}=\partial_{\mu}J_{L}^{\mu}=\frac{N_{f}}{32\pi^{2}}\left(g^{2}W_{\mu\nu}^{a}\tilde{W}^{a\mu\nu}-g'^{2}F_{\mu\nu}\widetilde{F}^{\mu\nu}\right)\label{e1}
\end{equation}
where $W_{\mu\nu}^{a}$ is the field tensor for the gauge group $SU\left(2\right)_{w}$
with gauge coupling $g$ and $F_{\mu\nu}$ is the field tensor
for the gauge group $U_{Y}$ with gauge coupling $g'$ . (The dual
tensors $\tilde{W}^{a\mu\nu}$and $\widetilde{F}^{\mu\nu}$ are defined
by $\tilde{W}^{a\mu\nu}=\frac{1}{2}\epsilon^{\mu\nu\alpha\beta}W_{\alpha\beta}^{a}$
and $\tilde{F}^{\mu\nu}=\frac{1}{2}\epsilon^{\mu\nu\alpha\beta}F_{\alpha\beta}$.)
From Eqn.(\ref{e1}) it follows that $B-L$ is an exact symmetry of the quantum theory, while $B+L$ is anomalous. Hence, in this framework,
non-conservation of $L$ implies non-conservation of $B$; so leptogenesis
 implies baryogenesis. It was shown in \cite{'t Hooft} that processes which violate $B+L$ correspond to transitions between inequivalent gauge-field vacua, known as instantons \cite{Jackiw Rebbi, Callan Dashen Gross}. However, the probability of tunnelling is suppressed by an exponential factor governed by the potential barrier between vacua. 
The potential barrier can be overcome at non-zero temperature \cite{Kuzmin-Rubakov-Shaposhnikov}.
 This scenario, where leptogenesis implies baryogenesis, holds clearly within models in which
SM  can be  embedded. 
\item Invariance with respect to $C$ is manifestly broken in SM; invariance with
respect to $CP$ is broken by complex phases in the the Yukawa couplings. 
\item The expansion of the Universe provides out-of-equilibrium conditions.
A first order electroweak phase transition can provide a non-equilibrium
situation at the transition temperature. However, from the observed value of the Higgs mass, the transition is predicted to be continuous and, for this reason, it cannot lead to a significant departure from equilibrium \cite{Buchmuller electroweak, Kajantie electroweak}.
\end{itemize}
The SM, although it satisfies the Sakharov conditions, leads to a prediction for $Y_{\triangle B}$ which is several orders of magnitude smaller than its observed value \cite{Kuzmin-Rubakov-Shaposhnikov}.   
Extra sources of $CP$ violation beyond SM are needed. The mechanism proposed in this paper provides a new universal source of $CP$ violation from a gravitational background arising in string theory. Moreover this same background breaks $CPT$ invariance.  

We shall consider leptogenesis as the path to baryogenesis following the important scenario for leptogenesis pioneered by Fukugita and Yanagida~\cite{Fukugita-Yanagida, Luty}: the lepton abundance is produced by the decay of heavy right-handed Majorana neutrinos (and so represents physics beyond the Standard Model (BSM)). The difference in the branching ratios of the channels of production of leptons and antileptons is equal to the imaginary part of the interference term of tree-level and one-loop diagrams for the decay processes. For the interference to generate a non-zero CP violating phase, at least two 
generations of right-handed neutrinos are needed (see \cite{Fukugita-Yanagida} and formulae therein).\footnote{ A pedagogical discussion of the necessity  (in the absence of torsion) of  interference between tree-level and one-loop diagrams can be found in \cite{Chen:2007fv}.}
Three right-handed neutrinos are required in the see-saw mechanism~\cite{Seesaw} 
for the generation of light neutrino masses; the observed phenomena of 
neutrino flavour oscillations among the light neutrinos requires neutrinos to have mass.
Measurements on solar, atmospheric and reactor neutrinos have established that
there are oscillations with distance of neutrino flavours \cite{Neutrino Oscillation}. The model of Fukugita and Yanagida 
connects an explanation of leptogenesis with the see-saw mechanism. The model thus represents 
an economical extension of SM. The fermions occurring in this leptogenesis model will couple to the gravitational background; this will have interesting consequences.

String theory~\cite{string} provides a framework which includes gravity. Although there are different types of string theory, the low energy actions that emerge contain the massless fields: graviton, a scalar field, the dilaton, and a spin-one antisymmetric tensor field $\mbox{\frakfamily{B}}_{\mu\nu}=-\mbox{\frakfamily{B}}_{\nu\mu}$, the Kalb-Ramond field~\cite{Kalb:1974yc}. The geometry due to a background  Kalb-Ramond field can lead to a Lorentz and CP violating interaction with fermions  \cite{Mavromatos-Sarkar,Environmental}. The corresponding field strength $H_{\mu\nu\rho} = \partial_{[\mu}\, \mbox{\frakfamily{B}}_{\nu\rho]}$ (where $[\dots ]$ denotes antisymmetrisation of the respective indices) plays the r\^ole of torsion of the background geometry, and is universally coupled to fermions via the affine connection.
Such couplings (in specific backgrounds) belong to the class of interactions considered in the extension of the SM (SME) proposed in \cite{Colladay-Kostelecky} and can be
both Lorentz, CP and CPT violating. Moreover, in four space-time dimensions, it follows from the equations of motion that the dual of the $H$ field strength, $\epsilon^{\mu\nu\rho\sigma} H_{\mu\nu\rho}$ may be represented as $\partial^\sigma b(x)$, where $b(x)$ is a pseudo scalar field - the `Kalb-Ramond' (KR) axion. 

\begin{figure}
\includegraphics[width=0.4\columnwidth]{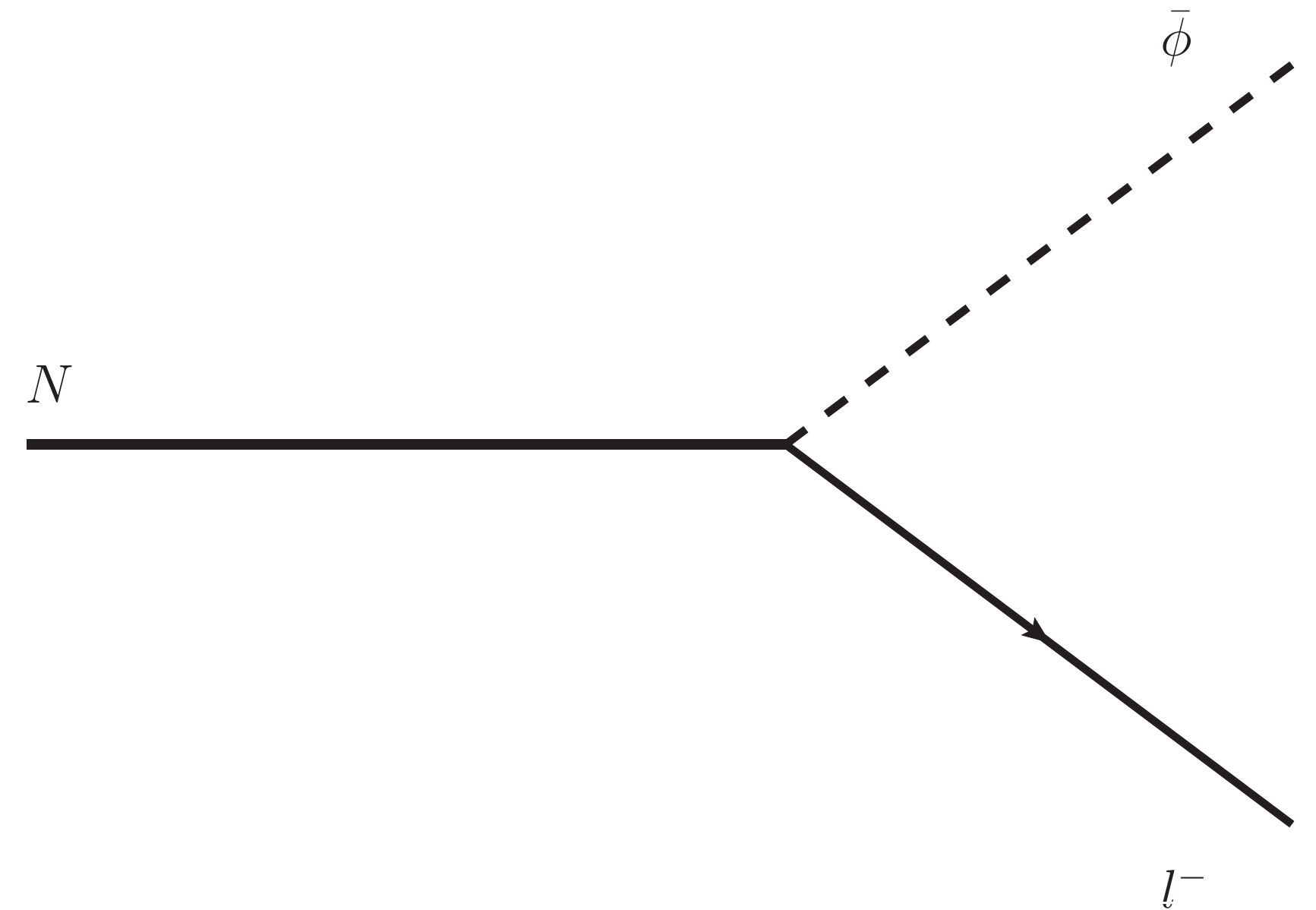}
\includegraphics[width=0.4\columnwidth]{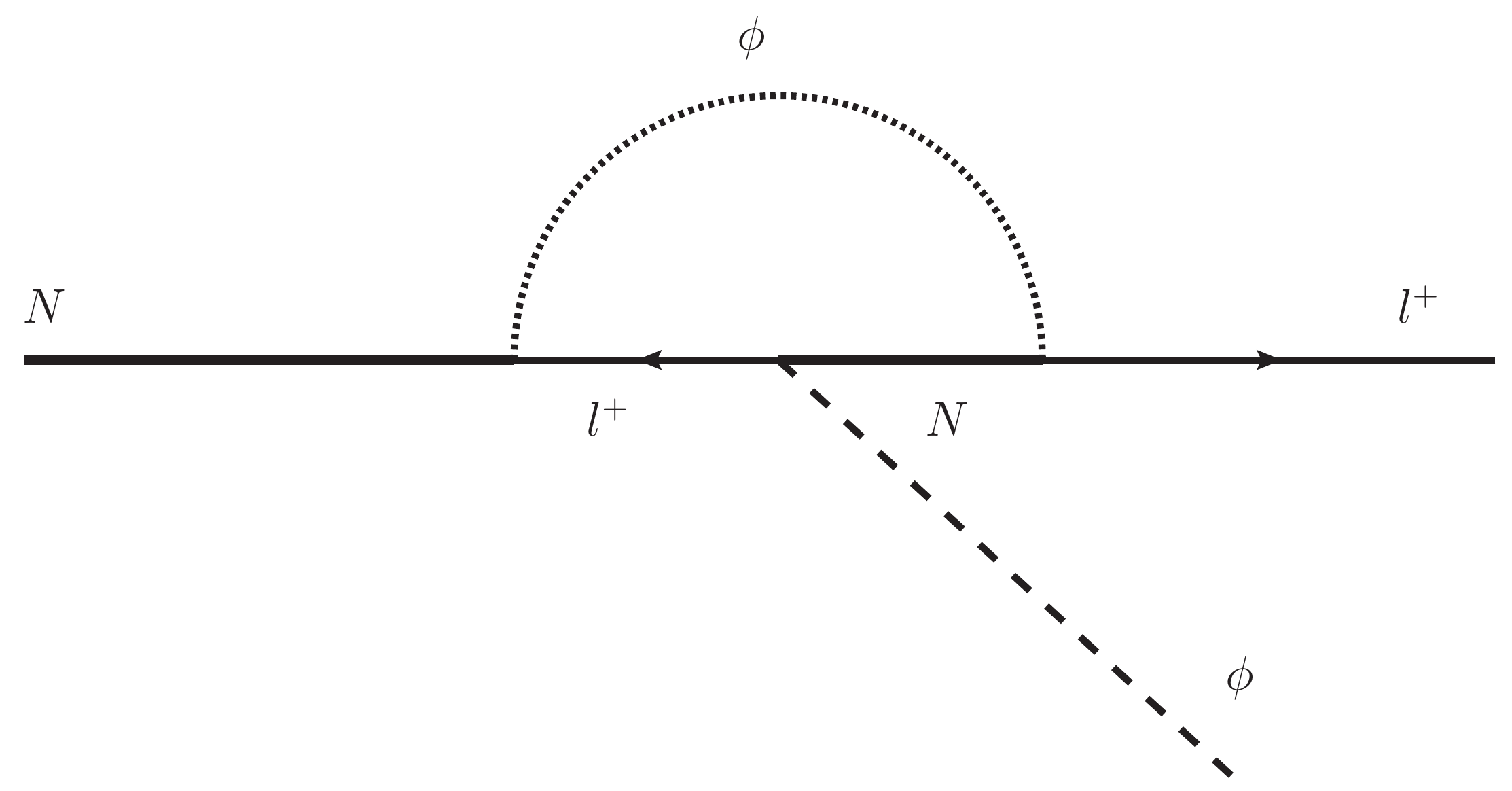} 
\caption{Tree- (left) and one-loop (right) decay amplitudes, corresponding to the Yukawa term that couples a right-handed neutrino to the standard model lepton sector. Analogous diagrams describe the decay in anti-leptons. Continuous undirected lines represent right-handed neutrinos, lines with an arrow are used to represent SM leptons,  whilst dashed lines correspond to the SM Higgs. The left diagrams are understood to be evaluated in the presence of a KR background field. The right diagram is the standard result of \cite{Fukugita-Yanagida}, leading to Leptogenesis.}
\label{fig:tree}
\end{figure}

If the KR torsion field was large enough in the early Universe, we will show that sufficient leptogenesis can occur even with only one  right-handed neutrino. A further feature is that the lepton asymmetry can be obtained even by only considering tree level Feynman diagrams. 
The diagrams represent the decays of a right-handed neutrino to a Higgs particle and a light left handed lepton or the corresponding anti-lepton (because of Yukawa couplings). In order to study consistently such decays, the 
external lines of the pertinent Feynman diagrams must be treated non-perturbatively in the external field strength. When more generations of right-handed neutrinos are considered, there is an additive tree-level modification to the standard (one-loop) expression of the asymmetry derived in \cite{Fukugita-Yanagida} (\emph{cf}. fig. \ref{fig:tree}). 
 On embedding our theory into the type 1 see-saw models, we would naturally consider three right handed neutrinos. However, if the masses of these heavy right-handed neutrinos are hierarchical, our  considerations for leptogenesis would reduce to considering the lightest of these right-handed neutrinos\footnote{There are scenarios \cite{MavPilaftsis}, in which a KR axion mixes (via the respective kinetic terms) with ordinary axion fields, which in turn couple via appropriate chirality changing Yukawa couplings to massless chiral neutrino fields. Such Yukawa couplings generate Majorana masses for the chiral neutrino fields through higher loop anomalous graphs, involving graviton fields without the need to have any specific number of right-handed neutrinos. We could then embed our mechanism for leptogenesis into a scenario with just one right-handed neutrino but with masses for all the left-handed neutrinos.}.

In the next section \ref{sec:KR}, we shall review the formalism underlying the Kalb-Ramond torsion $H$-field and discuss under what conditions the coupling of fermions to the $H$-field resembles that of a Lorentz and CPT violating coupling of fermions in an axial constant background field in the SME. In section \ref{sec:prop} we discuss dispersion relations, spinor chirality and helicity properties of Majorana spinors in such backgrounds. These  properties will be useful when we discuss in section \ref{sec:model} our model, an extension of the standard model with one right-handed neutrino coupled to the standard model sector via appropriate Yukawa couplings, formulated in a space-time with torsion, represented by a constant axial background field in the observer's frame.  We emphasise in section \ref{sec:lepto} the r\^ole, as sources of leptogenesis, of the CP-violating tree-level decays of heavy right-handed neutrinos, depicted in fig.~\ref{fig:tree},  in the presence of the axial background. Finally conclusions and outlook are presented in section \ref{sec:concl}. Some technical aspects of our work are given in three Appendices.

\section{Kalb-Ramond torsion \label{sec:KR}}

In general relativity gravity is represented by the curvature of spacetime. The connection on spacetime is taken to be torsion-free and metric-compatible. Hence it is uniquely determined to be the Levi-Civita connection once the metric is given. More generally we  have two independent 1-forms
\begin{equation}
e^{a}\equiv e^{a}_{\mu }\left ( x \right )dx^{\mu },
\qquad \omega _{b}^{a} \equiv \omega _{b\mu }^{a}\left ( x \right )dx^{\mu }.
\end{equation}
We can introduce two 2-forms: the curvature 2-form 
$R_{b}^{a}=d\omega_{b}^{a}+\omega_{c}^{a}\wedge\omega_{b}^{c}$ and the torsion 2-form $T^{a}=de^{a}+\omega_{b}^{a}\wedge e^{b}$. If $T^{a}$ vanishes then $\omega_{b}^{a}$ and $e^{a}$ 
are not independent.  From the principle of general covariance we know that we have an $SO(3,1)$ local invariance (this is manifest in the tetrad formalism). We can go from Lorentz and space-time indices via 
 \be
\gamma^{\mu}\left(x\right)=e_{a}^{\mu}\left(x\right)\gamma^{a},\qquad
g_{\mu\nu}\left(x\right)=\eta_{ab}e_{\mu}^{a}\left(x\right)e_{\nu}^{b}\left(x\right),\qquad
e_{a}^{\mu}e_{b\mu}=\eta_{ab}
 \ee 
 where $\gamma^{\mu}\left(x\right)$  is the Dirac matrix , $g_{\mu\nu}\left(x\right)$ is the metric  and $\eta_{ab}$ is the Minkowski metric.
 
The torsion~\cite{kibble,Shapiro}, in terms of space-time indices, is a rank $\begin{pmatrix} 1 \\ 2 \end{pmatrix}$  tensor, antisymmetric in the lower indices $T^{\lambda}_{\mu\nu}=-T^{\lambda}_{\nu\mu}$. No clear evidence exists  for a classical torsion field. Nevertheless, there have been several results on torsion phenomenology which can be found in the published literature (see for example \cite{Shapiro,poplawski,ellis,magueijo}). However there is at least one good (theoretical) reason to allow the spacetime connection to have a non-vanishing torsion: the gravitational multiplet of string theory~\cite{schwarz}.

\subsection{String-Theory-Induced Geometry with Torsion }

In string theory~\cite{string} the gravitational multiplet  includes the graviton, described by the metric tensor $g_{\mu\nu}$, the   (scalar) dilaton $\phi$ and the anti-symmetric Kalb-Ramond field $\mbox{\frakfamily{B}}_{\mu\nu}$. To first order in the string amplitude, the bosonic part of the low-energy effective action in the gravitational sector is given by \cite{schwarz}
\be\label{low energy effective action}
S=\frac{1}{2\kappa^2}\int \mbox{d}^4x\;\sqrt{-g}\left(R-2\partial^{\mu}\phi\partial_{\mu}\phi-e^{-4\phi}H_{\lambda\mu\nu}H^{\lambda\mu\nu}-\frac{2}{3}\delta c\exp\left(2\phi\right)\right),
\ee
where $\frac{1}{\kappa^2} \equiv \frac{M_{s}^2V^{c}}{8\pi} = \frac{1}{8\pi {\rm G}}$, with ${\rm G}$ the four-dimensional (gravitational) constant, $M_{s}^2$ the string mass scale, $V^{c}$ the (dimensionless) compactification volume in units of the Regge slope $\alpha^\prime$ of the string and $\delta c$ is the central charge deficit. It is possible, using a construction with a brane and associated bulk, to arrange  for a negative  $\delta c$~\cite{rizos}. Since $\phi$ varies slowly the deficit term behaves  like a negative contribution to the cosmological constant. Thus there is a close parallel with the Randall-Sundrum brane world picture~\cite{randal} which also has an anti-deSitter metric in the bulk.The field $H_{\lambda\mu\nu}$ appearing in the formula represents the field strength of the Kalb-Ramond field and is defined in analogy with the electromagnetic tensor $H_{\lambda\mu\nu}=\partial_{[\lambda}\mbox{\frakfamily{B}}_{\mu\nu]}$. Square brackets denote antisymmetrization over the enclosed indices. It can be shown that the sum of the graviton and the Kalb-Ramond terms in (\ref{low energy effective action}) can be re-written as the scalar curvature $\overline{R}$ of a new connection \cite{schwarz} defined as 
 
\be\label{connection}
\overline{\Gamma}^{\lambda}_{\;\mu\nu}=\Gamma^{\lambda}_{\;\mu\nu}+e^{-2\phi}H^{\lambda}_{\;\mu\nu} \ne \overline{\Gamma}^{\lambda}_{\;\nu\mu}.
\ee
In the string effective action this can be extended to include corrections~\cite{sloan,kaloper} of higher order in $\alpha^\prime$ .  The antisymmetry of $H^\lambda_{\;\mu\nu}$ in its lower indices, implies that this field strength plays the r\^ole of a torsion tensor~\cite{kibble,Shapiro}.
This suggests that this new connection (\ref{connection}) might be more fundamental than the Levi-Civita connection, and leads to different predictions whenever the Kalb-Ramond field is in a non-trivial configuration. In \cite{Mavromatos-Sarkar,Environmental} a potential r\^ole of the $H$ field for leptogenesis was emphasised. It is the purpose of this article to elaborate further on this issue.

The connection in (\ref{connection}) allows one to formulate the dynamics of matter fields minimally coupled to the gravitational and torsion background. In particular, the case of a Dirac spinor will be considered. Non-minimal couplings of matter fields to torsion have also been considered  in \cite{Shapiro}. The definition of the covariant derivative of a spinor requires the introduction of the tetrad $\{e_{a}^{\mu}\}$.

In the local Lorentzian frame given by the tetrad, the action is the same as the flat one, provided that ordinary derivatives are replaced by covariant ones $\partial_{a}\rightarrow\widebar \nabla_{a}$ (minimal coupling). This is obtained by requiring that $\widebar \nabla_{a}\psi$ transforms under a boost of the tetrad according to the spinor and vector indices it carries~\cite{kibble}. The result that one finds in this way is the following
\be
\widebar \nabla_{a}\psi=e_{a}^{\mu}\left(\partial_{\mu}+\frac{i}{2}\widebar \omega_{b\mu c}\Sigma^{bc}\right)\psi.
\ee
In the formula above $\Sigma^{ab}=\frac{i}{4}\left[\gamma^{a},\gamma^{b}\right]$ is the generator of the Lorentz group representation on four-spinors, while $\omega_{a\mu b}$ is the Ricci rotation coefficient, defined as
\be\label{omegabar}
\widebar \omega^{a\; b}_{\, \mu}=e^{a}_{\nu}\widebar\nabla_{\mu}e^{b\nu}=e^{a}_{\nu}\left(\partial_{\mu}e^{b\nu}+\overline{\Gamma}^{\nu}_{\mu\lambda}e^{b\lambda}\right).
\ee
Therefore the action is
\be
\label{diracac}
S_{Dirac}=\frac{1}{2} \int  \mbox{d}^4x\;\sqrt{-g}\, i\left(\overline{\psi}\gamma^{a}\widebar \nabla_{a}\psi-\widebar \nabla_{a}\overline{\psi}\gamma^{a}\psi 
+2im\overline{\psi}\psi\right).
\ee
The second term is usually not written in flat space, as its contribution is equal to the first term plus a surface integral. However, the situation is radically different when spacetime is not flat. In fact the second term is needed in order preserve unitarity, allowing for the cancellation of an anti-hermitean term involving the trace of the Ricci coefficients $\omega^{a}_{\;ac}$.

At this point we would like to clarify the physical content of the new terms contained in the spin connection. Using the gamma matrix algebra $\{\gamma^{a},\gamma^{b}\}=2\eta^{ab}$ and the definition $\gamma^5=i\gamma^0\gamma^1\gamma^2\gamma^3$, it turns out that the Dirac action (\ref{diracac}) 
can be rewritten in the following way
\be\label{diracb}
S_{Dirac}=\int  \mbox{d}^4x\;\sqrt{-g}\, \overline{\psi}\left(i\gamma^{a}\partial_{a}+\widehat B_{d}\gamma^5\gamma^d-m\right)\psi \equiv S_{Dirac}^{\rm free} + \int d^4x \sqrt{-g}\, \widehat B_\mu J^{5 \, \mu}~, \quad J^{5 \, \mu} \equiv \widebar \psi \, \gamma^\mu \, \gamma^5 \, \psi\ee
The axial vector $\widehat B^{d}$ is defined by
\be\label{bddef}
\widehat B^{d}=\frac{1}{4}\varepsilon^{abcd}e_{a}^{\;\mu} \widebar \omega_{b\mu c} = \frac{1}{4}\varepsilon^{abcd}e_{a}^{\;\mu}\, e_{b\nu}\left(\partial_{\mu}e_c^{\, \nu} + e^{-2\phi}H^{\nu}_{\;\mu\lambda} \, e_c^{\, \lambda}\right)
\ee
where in the last step we used  (\ref{omegabar}), (\ref{connection}) and the symmetry properties of the torsion-free Christoffel symbol $\Gamma^\lambda_{\mu\nu} = \Gamma^\lambda_{\nu\mu}$. 
 Owing to quantum fluctuations of the background, there will be an important additional contribution to $\widehat B^{d}$ from the fermion axial current. This will be discussed in a future section and in particular see Eq.~(\ref{hdel}).

 In the special case of either flat Minkowski or space times which do not contain off-diagonal metric elements mixing temporal and spatial components, 
such as Robertson-Walker Universes, of interest to us here, 
the axial vector $B^{d}$ is non-trivial and constitutes just the dual of the torsion tensor 
\be\label{defB}
\widehat B^d = -\frac{1}{4} \, \varepsilon^{abcd}\, e^{-2\phi}\, H_{abc}~.
\ee
In four space-time dimensions 
\begin{equation}\label{bfield}
\widehat B^\mu = \partial^\mu b~,
\end{equation}  
where $b(x)$ is a pseudoscalar field (also termed the Kalb-Ramond (KR) axion field).

However, on a generic spacetime there is also a derivative coupling of the spinor to the tetrad. Notice that this effective interaction with the gravitational background is not the only intricacy in dealing with spinors in curved space, as the kinetic term itself involves the tetrad $\partial_{a}\equiv e_{a}^{\mu}\partial_{\mu}$ and is therefore dependent on the space-time point. The important point here is that Dirac spinors are naturally coupled to an axial field, whose origin is purely gravitational (i.e. derives from the gravitational multiplet of string theory). We will show that this interaction leads to interesting cosmological consequences.

For a  string theory (with four uncompactified dimensions)  in non-trivial cosmological backgrounds, a world-sheet description is provided by a sigma model that can be identified with a Wess-Zumino-Witten type conformal field theory~\cite{Ellis-Nanopoulos-etc}.This construction has led to exact solutions (valid to all orders in the Regge slope, $\alpha^\prime$) for cosmological  bosonic backgrounds  with non-trivial metric, antisymmetric tensor and dilaton fields. Such solutions, in the Einstein frame, consist of (i) a Robertson-Walker metric with a scale factor  $a(t) \sim t$  where $t$ is the cosmic time, (ii) a dilaton field  $\phi$ that scales as   $\phi (t) \sim -{\rm ln}a(t), $ and (ii) a KR axion field scaling linearly with the cosmic time, $\overline {b} \propto t$ with $\overline b$ denoting the background value of $b$ (\emph{cf}. (\ref{bfield}). The resulting background axial vector $\overline{{\widehat B}}^d$ has only a non-trivial temporal component         
\begin{equation}\label{cosmolsol}
\overline b \sim {\rm const}\,  t ~, \qquad \partial^\mu \overline b \sim \epsilon^{\mu\nu\rho\sigma}\, e^{-2\phi} \, H_{\nu\rho\sigma}~,
\end{equation}and
\begin{equation}\label{constbdot}
\overline{{\widehat B}}^0 \propto {\dot {\overline b}} = {\rm constant}  
\end{equation}
in the Robertson-Walker frame. Calculations of string amplitudes involving fermions imply that the above gravitational background characterises the effective action in the presence of fermions (at least in lowest order in $\alpha^\prime$). Lorentz invariance does not hold, in the sense that it is spontaneously broken by the vacuum. The early-universe eras are also characterised by a high temperature and high density of relativistic matter. Consequently there can be non-vanishing components of vacuum currents. The requirement of maintaining rotational symmetries, implies that the temporal components of currents are allowed to condense. 
The low-energy string effective action for four-dimensional graviton and antisymmetric tensor backgrounds, 
 in the presence of fermions coupled to the torsion $H$-field as in (\ref{diracb}), gives the following equations of motion :
\begin{eqnarray}\label{eqsmot}
&& {\rm graviton:}  \qquad R_{\mu\nu} - \frac{1}{4} H_{\mu}^{\,\, \alpha\beta} H_{\nu\alpha\beta} = 8\pi G \, T^\psi_{\mu\nu} - \frac{1}{2} g_{\mu\nu} T^\psi + {\rm dilaton-derivative~ terms}  +\dots ~,\nonumber \\
&& {\rm antisymmetric~tensor:}  \qquad  \partial^\mu \Big(\sqrt{-g} e^{-2\phi}\,  \big[ H_{\mu\nu\rho} - \epsilon_{\mu\nu\rho\sigma} J^{5\, \sigma} + \dots \,\big] \Big) = 0~,
\end{eqnarray}
where $\dots$ denotes higher order terms in $\alpha^\prime$, $T^\psi_{\mu\nu} $ is the stress-energy tensor of fermionic matter and $T^\psi = g^{\mu\nu} T_{\mu\nu}^\psi$. There is of course an equation of motion for the dilaton which provides additional constraints for the background. 

From the equation for the antisymmetric tensor field, assuming a constant dilaton for simplicity, we observe that it can be solved upon using the pseudoscalar dual field $\overline b$ defined in (\ref{cosmolsol}): 
\begin{equation}\label{jb}
\partial^\mu \Big(\sqrt{-g} \big[ \epsilon_{\mu\nu\rho\sigma} (\partial^\sigma {\overline b} -  \tilde c \,  J^{5\, \sigma}) + \mathcal{O}\Big((\partial \overline b)^3\Big) \,\big] \Big) = 0~,
\end{equation}
where $\tilde c$ is a constant of proportionality, and we replaced $\dots$ in (\ref{eqsmot}) by their explicit form in terms of the KR axion.  In the absence of fermionic currents, it is these higher order terms (resummed to all orders in $\alpha^\prime$) that are responsible for the exact solution  (\ref{constbdot}), which is derived in \cite{Ellis-Nanopoulos-etc} using conformal field theory.
The existence of a non-trivial Lorentz-violating temporal component of the fermion vacuum current, expected at high-temperatures and densities of the early-universe era, is consistent with a constant $\partial_t \overline b \equiv \dot{\overline{b}}$ of the gravitational sector, even at lowest order in $\alpha^\prime$ (which we assume characterises the leptogenesis era)~\footnote{Of course, in a Robertson-Walker space-time, there is also another solution to lowest order in $\alpha^\prime$ scaling with the scale factor of the Universe $a(t)$ as $\dot{\overline b} \sim a^{-3}$. In view of the landscape  
nature of string theory the different solutions are  possible, depending on the energy scale.}. The value of this constant is of course different between the two solutions. This follows from (\ref{jb}) since 
\begin{equation}\label{constbdot2}
   \dot{\overline{b}} = \tilde c\, \langle J^{5}_0 \rangle =\tilde c  \, \langle \psi^\dagger_i \gamma^5 \psi_i \rangle  = {\rm constant} \ne 0 ~,
\end{equation}
where $i$ runs over appropriate fermion species that are allowed to condense in the early Universe. From, the graviton equation of motion, then, it can be readly seen that, in the context of a Friedmann-Robertson-Walker Universe, the r\^ole of the time dependent pseudoscalar, with constant rate (\ref{constbdot2}) is to provide a vacuum energy term of the type of a positive cosmological constant $R_{\mu\nu}  = g_{\mu\nu} (\dot{\overline{b}})^2  +  8\pi G \, T^\psi_{\mu\nu} $. However we have a negative contribution due to the central charge deficit term. The balance between these contributions is arranged in order to have cosmological stability.

We assume that our bosonic stringy backgrounds with constant torsion are non-thermal~\cite{Environmental}, 
and, as discussed above, characterise more complicated, phenomenologically realistic string-inspired cosmologies;
one may then study the background-induced CP-violation effects on lepton asymmetry. This will be the main point of the article. 
If the Universe undergoes a phase transition at a temperature $T_c $ (after the leptogenesis era), where the 
axial current condensate vanishes, then, from (\ref{jb}) we obtain immediately that, for a Robertson-Walker Universe, with a scale factor $a(t)$, 
upon ignoring (as subleading) the higher order $\mathcal{O}\Big((\partial \overline b)^3\Big)$ terms,  
the rate of change of the $\overline{b}$ field diminishes with the cosmic time as the cube of the scale factor
\begin{equation}\label{bscale}
\dot{\overline{b}} \sim 1/a^{3}(t)~. 
\end{equation}
From the point of view of the underlying conformal field theory, the phase transition (or a series thereof) is assumed to be such so as to reduce the corresponding central charge~\cite{Ellis-Nanopoulos-etc}.
We shall make use of this result when we discuss the history of this Universe after the leptogenesis epoch in section \ref{sec:abund}. 

Before that we would like to discuss the effects of quantum fluctuations of the torsion. 
It can be seen from the effective string action (\ref{low energy effective action}), at lowest-order in $\alpha^\prime$, that the $H$-field is non-propagating; hence it would seem that it can be integrated out exactly in a path integral. However, 
in the context of a string-inspired low-energy theory effective action, integrating out the torsion $H$-field  is non-trivial: 
the action contains an infinity of higher-derivative interactions, among them those containing $(\nabla H)^2 $ terms~\cite{sloan}, that make the $H$-field a full fledged propagating field with complicated interactions, of infinite order in $\alpha^\prime$ (which are not known in closed form).
In the context of this work we shall assume weakly varying $H$-fields, where one can restrict oneself to the lowest-order effective action (\ref{low energy effective action}) where the $H$-torsion can be integrated out, as a non-propagating field, mirroring the case of ordinary torsion in Einstein-Cartan theory, reviewed in Appendix \ref{appendix1}. 
In the next section we will discuss  quantum aspects of torsion in the truncated theory. As will be shown, the main effect of torsion (in this approximation) is to introduce repulsive axial current-current interactions among the fermions.
We shall also derive consistency conditions for the existence of constant torsion backgrounds in the presence of fermions. Consistency is achieved by the formation of a condensate of the temporal component of the axial current.

\subsection{Quantum Fluctuations of Torsion and Four Fermi interactions}\label{pathintegral}

An important aspect of the effective action for torsion is the absence of a kinetic energy term for the torsion. As a consequence, in the path integral approach to quantum theory, we can integrate out  the antisymmetric torsion field strength $H_{\mu\nu\rho}$. This feature helps to link our effective  theory to the more general case of fermions in spacetimes with generic torsion (see Appendix \ref{appendix1}). In string theory, $\mathbf{H}$ is not typically an exact form since anomaly cancellation requires that the antisymmetric tensor field strength is augmented with (anomalous) gravitational and gauge Chern-Simons  terms~\cite{string}
\be
\label{gdh}
\mathbf{H} = \textbf{d} \mathbf{B}+ \frac{\alpha^\prime}{4} (\boldsymbol{\omega }_{3L} - \boldsymbol{\omega }_{3Y} )
\ee
where $\alpha^\prime$ is the Regge slope of the string;
$\boldsymbol{\omega}_{3L}$ is the Lorentz Chern-Simons term associated with the spin connection $\boldsymbol{\omega}$;
$\boldsymbol{\omega}_{3L} =   {\rm Tr}\Big[\boldsymbol{\omega} \wedge ( d\boldsymbol{\omega} + \frac{2}{3} \, \boldsymbol{\omega} \wedge \boldsymbol{\omega}) \Big]$ 
and $\boldsymbol{\omega}_{3Y}$ is the (non-Abelian in general) gauge field; 
$\boldsymbol{\omega}_{3Y} = {\rm Tr} \Big[\mathbf{A}\, \wedge (d \mathbf{A}  + \frac{2}{3} \mathbf{A} \wedge \mathbf{A})\Big]$. 
We have been schematic  rather than precise in the expressions of the various Chern-Simons forms; the precise form of the terms is not directly relevant for us. 
Within our four dimensional space time setting, the various Chern-Simons forms appear as the compactified form of the corresponding ten-dimensional ones that exist in the  higher dimensional effective theory. 

The generalisation (\ref{gdh})  leads to the following modified Bianchi identity for the generalised field strength~\cite{kaloper}
\be\label{bianchi} 
d \mathbf{H} = \frac{\alpha^\prime}{4} \Big({\rm Tr}\mathbf{R} \wedge \mathbf{R} - {\rm Tr} \textbf{F} \wedge \textbf{F}\Big)
\ee
where $\mathbf{F} $ is the gauge field strength and $\mathbf{R}$ is the Riemann curvature form, with respect to the torsion-free connection. 
In terms of the dual of the $H$-field, $\widehat B^\mu$ field (\ref{defB}), the Bianchi identity can be written  in components as:
\be\label{bianchiB}
\nabla_{\mu} \, \widehat B^\mu = \frac{1}{32}\alpha^\prime \Big(R_{\mu\nu\rho\sigma} \, {\widetilde{R}}^{\mu\nu\rho\sigma} - F_{\mu\nu} \, \widetilde{F}^{\mu\nu} \Big) \equiv \alpha^\prime {\mathcal G}(\omega, \mathbf{A})~,
\ee 
where $\nabla_{\mu}$ denotes covariant derivative with respect the torsion-free connection, and $\widetilde{F}_{\mu\nu} = \frac{1}{2} \sqrt{-g}\, \varepsilon_{\mu\nu\rho\sigma} \, F^{\rho\sigma} $,  
 $\widetilde{R}_{\alpha\beta\mu\nu} = \frac{1}{2} \sqrt{-g} \, \varepsilon_{\mu\nu\rho\sigma} \, R_{\alpha\beta}^{\,\,\,\,\,\,\,\,\rho\sigma} $ are the corresponding dual tensors in four space-time dimensions.

The Bianchi identity (\ref{bianchiB}) can be implemented in the (Minkowskian signature space-time) path integral through a delta function, which, in turn, can be expressed as an integral over a Lagrange multiplier \emph{pseudoscalar} field 
$\sqrt{ \frac{8}{3 \,\kappa^2}}\, b$ 
(the normalisation has been chosen so that the $b$-field has a canonical kinetic term). In the context of the lowest order in $\alpha^\prime$ action (\ref{low energy effective action}), the $H^{\mu\nu\rho}$ (or equivalently the axial vector field) is non-propagating and so it can be integrated out: the integration is done by Euclideanising the path integral, after the introduction of the Lagrange multiplier field, performing the $B$-integration, and then by switching back to the Minkowski signature by an appropriate analytic continuation.
We split the $\widehat B$ field into background $\overline{\widehat{B}}^\mu $ and quantum fluctuations, ${\widehat B}^\mu$, 
\begin{equation}
B^\mu = \overline{\widehat{B}}^\mu + \widehat{B}^\mu 
\end{equation}
where the background satisfies (\ref{constbdot}). 
The result for the relevant factor of the path integral after integration over the quantum fluctuations $\widehat B$ reads  
\begin{eqnarray}\label{binteg}
\mathcal{Z} &\propto& \int D\psi\, D\overline\psi \, D\widehat B e^{iS (\overline{\widehat{B}} + \widehat B)+ 
iS_{Dirac}(\overline{\widehat{B}} + \widehat B)} \delta (\nabla_{\mu}\widehat B^\mu - \alpha^\prime{\mathcal G}(\omega, \mathbf{A}) ) = 
\nonumber \\
&=&  \int D\psi \, D\overline\psi  e^{i {\widetilde S}(\overline{\widehat{B}}) +  i\int d^4 x \, \sqrt{-g}\, \frac{3}{16}\, \kappa^2  J_\mu^5 \, J^{5 \, \mu}}\, 
\int \, db \, e^{i \int d^4 x \sqrt{-g}\, \Big[ \, \frac{1}{2} \partial_\mu b \, \partial^\mu b -  (\frac{3}{8})^{1/2}\,  \kappa\, J^{5 \, \mu}  \,  \partial_\mu b  -  \frac{\alpha^\prime}{\kappa}\,b\,{\mathcal G}(\omega, \mathbf{A}) \Big]}~, 
\end{eqnarray}
where $\widetilde S (\overline{\widehat{B}}) = S(\overline{\widehat{B}}) + S_{Dirac}^{\rm free} + \int d^4x \sqrt{-g}\, \overline{\widehat{B}}_\mu J^{5 \, \mu}$ is the background action, given by the sum of (\ref{low energy effective action}), and (\ref{diracb}). In arriving at the right-hand-side of (\ref{binteg}) we shifted the quantum field $b$ by the dual of the four-dimensional background 
$\overline b$, defined through (\ref{cosmolsol}), i.e. $b \rightarrow b +  \overline b$. In this way the shifted field contains only the quantum fluctuations. This operation decouples the background part of the action $\tilde S$ from the quantum fluctuation part, since it eliminates mixed terms of the form $\int d^4 \sqrt{-g} \overline{\widehat{B}}^\mu \, \widehat{B}_\mu $. 
Thus the presence of the non-propagating $H$-field in the quantum action leads to four fermion repulsive axial-current-current interactions~\footnote{In theories with anomalies~\cite{anomalies, Adler, Bell-Jackiw}, say U(1) axial anomaly, the axial current is not conserved but its covariant divergence is proportional to the term 
\begin{equation}\label{anomalies} 
\nabla_{\mu}J^{ 5 \, \mu} \propto {\mathcal G}(\omega, A)
\end{equation}
where A in such a case is a U(1) gauge field. 
However, insofar as quantum fluctuations of the torsion are concerned, which are presented by the path-integral factor over the field $b$ in (\ref{binteg}), one may add counterterms in the effective action such that the divergence of the (improved) axial current  in the torsion-fluctuation part of the effective action is treated as vanishing~\cite{kaloper,MavPilaftsis}. 
This implies that in (\ref{binteg}) the integral over $b$ field factorises, leaving the four-fermi axial-current interactions as the only effect of torsion quantum fluctuations. The fermions of this quantum-torsion theory have a four-fermion repulsive axial-current-current interaction, whose strength is weak, being proportional to the gravitational constant $\kappa^2 = 8\pi {\rm G}$. We shall come back to the role of such interactions in the next section. 
We would also like to make a comment on the consistency of the vacuum current condensates (\ref{constbdot2}) with the 
the anomaly equation (\ref{anomalies}) in a Robertson-Walker background, of interest to us. In such a case, by taking the vacuum expectation values on both sides of the equation, and assuming a Robertson-Walker space-time, we obtain
\begin{equation}\label{anomfrw}
 3 H \langle J^{5\, 0} \rangle \propto \langle {\mathcal G}(\omega, A) \rangle~, \,\langle J^{5\, 0} \rangle = {\rm constant}~.
\end{equation}
Thus, in view of the anomaly equation, the presence of such a condensate would be associated with a condensate involving gauge and gravitational fields, (\emph{cf.} (\ref{bianchiB})), which should scale with cosmic time as the Hubble parameter.  In string theory, whose effective action is characterised by highly non-linear terms in gauge field strength, such condensates are possible.}.

\subsection{Hehl-Datta non-linear spinor equations in the presence of torsion and particle-antiparticle asymmetries}

As we discussed above, in a spacetime with a non-vanishing torsion, 
the effective field theory of $N$ species of interacting (massive) fermions $\psi_i$ is given by the Lagrangian~\cite{Shapiro} 
(\ref{binteg}):
\be\label{fli}
\mathcal{L} e^{-1} =\frac{i}{2} \, e^\mu_a \Big( \overline{\psi}_j \gamma^a \psi_{j\, ;\, \mu} - \overline \psi_{j\, ;\, \mu } \gamma^a \, \psi_j \Big) + 
\overline \psi_j (\gamma^5 \, \slas{\widetilde{B}} -m^{(j)})\psi_j +  \frac{3\kappa^2}{16} (\overline \psi_j \gamma_\mu \gamma^5 \psi_j)\, 
(\overline \psi_\ell  \gamma^\mu \gamma^5 \psi_\ell) + \dots~,
\ee
where $e^\mu_a $ are the vierbeins, $e$  is the vierbein determinant, the suffix $;$ denotes covariant derivative with respect to the torsion-free space-time connection, 
$\widetilde {B}_\mu \equiv \overline{\widehat{B}}_\mu $, is the axial background,  and the $\dots$ denote other terms that may be present in the microscopic (string-inspired) theory. 
 Summation over the fermion flavour indices $j, \ell = 1, \dots N$ is understood. The non-renormalizable four-fermion interactions arise on integrating out the torsion field (as noted in (\ref{binteg})).

 We shall consider both Dirac and Majorana spinors in the framework of the interacting theory (\ref{fli}).
In order to discuss the effects of torsion on particle-antiparticle induced asymmetry, we commence this section with the equations of motion for the spinors (and their charge conjugates) that follow from the Lagrangian (\ref{fli}).

The four-fermion interaction  term  will  induce  a cubic  term  in  the
  equations  of motion  for  the fermions. Such non-linear equations first appeared in the 1970's work of Hehl and Datta~\cite{hehldatta} and are now known eponymously. Under  the assumption  of
  formation of a (Lorentz violating) fermionic condensate of the axial
  current, which thus linearises the Hehl-Datta equations, it was recently argued~\cite{poplawski} that Dirac fermions
  may   lead  to   C-  and   CPT-violating  differences   between  the
  fermion-antifermion   populations    in   the   finite   temperature
  environment  of the  Early Universe~\footnote{However, the author did not consider that, after integrating out the torsion, the effective four fermions interaction is such that all the fermionic species must contribute to one and the same condensate.}.
  
  We will consider the non-linear equations stemming from (\ref{fli}) for both the Dirac spinor and the charge-conjugate spinor $\psi^c = \mbox{C}{\overline \psi}^T$, where $T$ indicates matrix transposition, and C is the (unitary) charge conjugation matrix, $\mbox{C}=i \gamma^2 \gamma^0$, in standard notation (no sum over $j$ index):
\begin{eqnarray}\label{hde}
i \, e^\mu_{\, a} \gamma^a \psi_{j\,;\, \mu} - m^{(j)} \psi_j   + \gamma^5 \, \slas{\widetilde B} \, \psi  + \frac{3\, \kappa^2}{8} \, \Big(\overline \psi_\ell \,  \gamma^5 \, \gamma_a\, \psi_\ell \Big) \gamma^5 \gamma^a \, \psi_j &=& 0 \nonumber \\
i\, e^\mu_{\, a} \gamma^a \psi^c_{j\, ;\, \mu} - m^{(j)}  \psi_j^c   + \gamma^5 \, \slas{\widetilde B}\, \psi_j^c  - 
\frac{3\, \kappa^2}{8} \, \Big(\overline \psi_\ell^c \gamma^5 \, \gamma_a \, \psi_\ell^c \Big) \gamma^5 \gamma^a \, \psi_j^c &=& 0~,
\end{eqnarray}
where, to obtain the second line, we used the Dirac equation obtained from (\ref{fli}) for the Dirac conjugate spinor, took the transpose ``\,$T$\,'', and acted upon from the left with 
 the C-conjugation operator, using   $-\mbox{C}\, \gamma_\mu^T \,\mbox{C}^{-1} = - \gamma_\mu $ and $\mbox{C}\, \gamma^{5\, T}\, \mbox{C}^{-1} = \gamma^5$. We also used
 that~\footnote{Notice that, since the Hehl-Datta equation is a classical equation, the object $\psi (x) $ represents a wave function in spinor space rather than a field operator. In this sense, to arrive at (\ref{conjdens}) only matrix transposition for fermion bilinears has been employed without changing sign, which would be the case if one dealt with second-quantised grassmann field operators $\psi$. In the latter case,  the axial current, of course, does not change sign under charge conjugation, as we shall see in subsequent sections.}
 \be\label{conjdens}
 \overline \psi_\ell ^c \gamma^5 \, \gamma_a \, \psi_\ell^c = - \overline \psi_\ell\,  \gamma^5 \,  \gamma_a \, \psi_\ell~.
 \ee 
  In a Hartree-Fock approximation, we may linearise the equations (\ref{hde}) by replacing the fermion bilinear in the non-linear terms with its vacuum expectation value $\mathcal{F}_{\mu}\equiv \langle\overline{\psi}\gamma^{5}\gamma_{\mu}\psi\rangle$.  For isotropic situations, as is the case we are interested in, only its temporal component is non trivial, and denotes the appropriate fermion chiral densities (summed up over all species). 
 \be\label{HF}
 {\mathcal F}_0  = \langle \psi_\ell ^\dagger \gamma^5 \psi_\ell \rangle \equiv \rho_R - \rho_L~, \qquad {\mathcal F}_i = 0~.
 \ee
 The linearised Hehl-Datta equations (\ref{hde}) become (assuming also only a $\widetilde B^0 \ne 0$ component, for concreteness, as appropriate for our string-inspired case (\ref{cosmolsol}), which we restrict ourselves to here)
 \begin{eqnarray}\label{hdel}
i \, e^\mu_{\, a} \gamma^a \psi_{j\,;\, \mu} - m \psi_j    - \Big( \widetilde B_0 + \frac{3\, \kappa^2}{8} \, {\mathcal F}_0 \Big)\, \gamma^0 \, \gamma^5 \, \psi_j &=& 0 \nonumber \\
i\, e^\mu_{\, a} \gamma^a \psi^c_{j\, ;\, \mu} - m \psi_j^c   - \Big( \widetilde B_0 +
\frac{3\, \kappa^2}{8} \, \mathcal{F}_0 \Big) \, \gamma^0\,  \gamma^5\, \psi_j^c &=& 0~.
\end{eqnarray} 
In \cite{poplawski}, the difference in sign of the cubic fermion terms in (\ref{hde}), between the fermions and their Dirac conjugate, have been interpeted as leading to different dispersion relations for constant background torsion and through this baryogenesis in the early Universe. 
Unfortunately we do not agree with this interpretation. In terms of (\ref{conjdens}) we observe that in a Hartree Fock approximation the isotropic condensate of the chiral current (interpreted as torsion) couples to matter and antimatter \emph{with the same sign}, and hence there is no induced difference in the corresponding dispersion relations.

 Moreover, a Majorana spinor can be defined as $\nu = \psi + \psi^c$ and is, by construction, a mass eigenstate, satisfying the Majorana condition  
$\nu^c  = \nu $, entailing that a Majorana fermion is its own antiparticle and is chargeless. From this condition, we observe that Majorana spinors do not contribute to the condensate ${\mathcal F}_0 $. Of course the torsion mixes the Majorana neutrinos with all other fermion species, and thus non trivial backgrounds ${\mathcal F}_0$ are experienced in general by Majorana fermions in such space times with torsion. It follows directly from the definition that a Majorana spinor satisfies the same equation (\ref{hdel}). We observe that the quantities ${\mathcal F}_0$ are in general temperature dependent. 

Motivated by the above discussion, in what follows we shall consider Majorana spinors in constant axial backgrounds, without specifying the microscopic nature of the background.

\section{Some properties of Spinors coupled to an axial background field \label{sec:prop}}
From now on we will consider the effective theory of spinors on flat space-time~\footnote{From now on we will use the metric signature $(-+++)$, which is most widely used in the particle physics community.} given by (\ref{diracb})
\be
S_{Dirac}=\int  \mbox{d}^4x\;\sqrt{-g}\, \overline{\psi}\left(i\gamma^{\mu}\partial_{\mu}+B_{\mu}\gamma^5\gamma^\mu-m\right)\psi
\ee
(To avoid cumbersome notation $B_{\mu}$ here will denote $ \Big( \widetilde B_{\mu} +
\frac{3\, \kappa^2}{8} \, \mathcal{F}_{\mu} \Big) $.)
This belongs to the class of theories termed Standard Model Extension (SME) and considered in \cite{Colladay-Kostelecky}. When the torsion is \emph{constant}  throughout spacetime the interaction term leads to the spontaneous breakdown of Lorentz symmetry. 
The interaction terms with coefficients $B_\mu$ are known to be \emph{both} Lorentz and CPT violating in that case.
This is in fact the case of the string-cosmology background (\ref{cosmolsol}) considered in \cite{Ellis-Nanopoulos-etc}, where only the $B_0$ (temporal) component is non-trivial. In the context of our present work we shall restrict ourselves mostly to this latter case.

\subsection{Dispersion Relations of Fermions in a constant axial background}

When $B_{\mu}$ is constant, it makes sense to look for plane wave solutions of the equations of motion. The dispersion relations thereby obtained are written in terms of the fourvectors $U_{\mu}=p_{\mu}-B_{\mu}$ and $V_{\mu}=p_{\mu}+B_{\mu}$ as
\be\label{general dispersion relation}
U_{\mu}U^{\mu}\;V_{\nu}V^{\nu}-2m^2 \; U_{\mu}V^{\mu}+m^4=0.
\ee
Hence for a fixed spatial momentum $\mathbf{p}$ there are four different values of the energy. Due to C-invariance of the operator $\overline{\psi}\gamma^5\gamma_\mu\psi$ the energy levels come in pairs with opposite sign. The relations found are the same that one would find when looking for the poles of the fermion propagator in \cite{Colladay-Kostelecky} and \cite{Alfaro}.

At this point we would like to observe that the dispersion relations in the paper \cite{Mukho}
$(p_a\pm B_a)=m^2$
(written in the tetrad frame in curved space, but supposedly valid also in flat space if one allows for torsion) are in general \emph{not} compatible with (\ref{general dispersion relation}) for massive fermions. In particular they are not in the case $\mathbf{B}=0$, $B_0\neq 0$. The peculiar form of  (\ref{general dispersion relation}) is essentially due to the chiral nature of the coupling. Because of the very definition of a fermion mass term, there is no natural way of splitting the two chiral components in the massive case and by no means they can be identified with particles and antiparticles. Also it is not clear under which conditions the field $B_a$ can be constant throughout on a curved spacetime, and even then how are plane wave solutions found in the non-holonomic basis. For a further discussion of this subject see \cite{Shore},  where the incorrectness of \cite{Mukho} is shown to follow from the non-tensorial transformation properties of the pseudo-vector $B_a$. The latter is a peculiar property of curved spacetimes, since on flat spacetimes the only contribution to the connection comes from the torsion tensor.  As such, the strong equivalence principle implies that, in Riemannian spaces without torsion,  locally one can always find a frame where the space-time is flat, thus eliminating $B_a$, which therefore cannot contain covariant information such as the one leading to leptogenesis. In contrast, the presence of a torsion field leads to a proper axial vector background coupled to fermions, which under certain circumstances may be constant in some frame, leading to CP and CPT violating Leptogenesis, \emph{e.g}. the case of stringy cosmologies~\cite{Ellis-Nanopoulos-etc}, 
in constant antisymmetric tensor field strength backgrounds in the Robertson-Walker frame.

From now on, unless otherwise specified, the constant field $B_{\mu}$ will be taken parallel to the time axis. This field value represents a vector v.e.v. that is responsible for the spontaneous breakdown of \emph{particle Lorentz invariance} (as defined in \cite{Colladay-Kostelecky}). A reason for this choice can be found in the papers \cite{Environmental} and \cite{Ellis-Nanopoulos-etc}, where solutions for the Kalb-Ramond field in the expanding universe are explicitly given. In fact, in that case $H_{\mu\nu\rho}=e^{2\phi}\varepsilon_{\mu\nu\rho\sigma}\partial^{\sigma}b(x)$ and the axion field $b(x)$ is linear in time, thus entailing a purely timelike $B_{\mu}$. The positive frequency spinors are given by
\be\label{positive-frequency solutions}
u_r(p)=\left(\begin{array}{c} \sqrt{E_r-B_0-\lambda_r|\vec{p}|} \;\xi^r\\ \sqrt{E_r+B_0+\lambda_r|\vec{p}|} \;\xi^r \end{array}\right)=\left(\begin{array}{c} \sqrt{E_r-B_0-\vec{p}\cdot\vec{\sigma}} \;\xi^r\\ \sqrt{E_r+B_0+\vec{p}\cdot\vec{\sigma}}\;\xi^r \end{array}\right).
\ee
$\xi^r(\vec{p})$ are the usual helicity eigenspinors
\be
\frac{\vec{p}\cdot\vec{\sigma}}{|\vec{p}|} \; \xi^r =  \lambda_r \; \xi^r, \quad r=1,2
\ee
where $\lambda_r\equiv(-1)^{r-1}$. The spinors are taken to be orthonormal, \emph{i.e.} $\xi^{r\dagger}\xi^{s}=\delta^{rs}$. $E_r$ is the energy corresponding to $\lambda_r$ via the relation
\be\label{dispersion relation}
E_r^2=m^2+(B_0+\lambda_r |\mathbf{p}|)^2.
\ee
The last equation is a particular case of (\ref{general dispersion relation}), corresponding to $\mathbf{B}=0$. This result agrees with the formulae in \cite{Colladay-Kostelecky,Adam Klinkhamer}.

On the other hand, \emph{negative-frequency} solutions are given by
\be
v_s(p)=\left(\begin{array}{c} \sqrt{E_s+B_0-\vec{p}\cdot\vec{\sigma}} \;\xi^s\\ -\sqrt{E_s-B_0+\vec{p}\cdot\vec{\sigma}} \;\xi^s \end{array}\right).
\ee
Notice that here the corresponding dispersion relations appear to be inverted if compared to the previous case. In other words, $E_1$ corresponds to $s=2$ and $E_2$ to $s=1$. This is  to be expected on the basis of Dirac's hole theory.

The spinors satisfy the following normalization conditions:
\be
\overline{u}_r u_s=2m\;\delta_{rs}, \quad \overline{v}_r v_s=-2m\;\delta_{rs},
\ee
or equivalently
\be
u_r^\dagger u_s=2E_r\;\delta_{rs}, \quad \overline{v}_r v_s=2E_r\;\delta_{rs}.
\ee

It is important to stress that these solutions only hold in the frame where $B_{\mu}$ is purely temporal. Furthermore, as a consequence of broken \emph{particle Lorentz invariance}, the spinor wave-function of a particle with momentum $\mathbf{p}$ cannot be obtained by boosting the solution for a spinor at rest.

The Dirac field operator is a straightforward generalization of the standard one and is constructed from the plane wave solutions of the Dirac equation \cite{Peskin}
\be\label{campo}
\psi(x)=\int\frac{\mbox{d}^3p}{(2\pi)^3}\sum_{r=1,2}\frac{1}{\sqrt{2E_r}}\left(a^{r}_{\mathbf{p}}u^r(p)e^{-ipx}+b^{r\dagger}_{\mathbf{p}}v^r(p)e^{ipx}\right).
\ee
 The only difference with the standard case is that here the energy depends on the helicity. As usual, canonical equal-time anti-commutation relations must be  imposed on the fields and this leads in turn to the fermionic oscillator algebra of the creation and annihilation operators $a^{r}_{\mathbf{p}}$, $a^{r \; \dagger}_{\mathbf{p}}$ and $b^{r}_{\mathbf{p}}$, $b^{r \; \dagger}_{\mathbf{p}}$.

\subsection{Inequivalence of helicity and chirality in the presence of an Axial Background}
It is well-known that, in the massless limit, the action of the chirality and the helicity operator on plane-wave solutions of the standard Dirac equation is the same. One can then ask whether this basic result still holds in the CPT violating theory considered. It turns out that the answer is negative, as it is readily seen with a simple example.

Let us consider a positive frequency, positive helicity spinor, as given by Eq. (\ref{positive-frequency solutions}), with $m=0$
\be
u_{1}=\left(\begin{array}{c} \sqrt{E_{1}-B_0-|\vec{p}|}\xi_{1}\\ \sqrt{E_{1}+B_0+|\vec{p}|} \xi_{1} \end{array}\right).
\ee
We have from (\ref{dispersion relation})
\be
E_{1}=\left\rvert B_0+|\vec{p}|\right\rvert.
\ee
When $B_0$ is non-zero, this leads to two different cases, depending on the sign of the argument of the absolute value function. Therefore, if $B_0+|\vec{p}|\geq0$ the spinor has right chirality, while $B_0+|\vec{p}|<0$ implies that it has left chirality! Obviously when $B_0=0$ the second case is forbidden, hence re-establishing the usual correspondence.

\section{Standard Model Extension with one right-handed neutrino in the presence of axial backgrounds \label{sec:model}}

Since the coupling of the axial field $B_\mu$ to the fermions is obtained via the minimal coupling prescription to the background torsion, it is universal \emph{i.e.} the coupling constant is the same for all fermionic species. Along with these CPT violating interactions, we further extend the SM by introducing (at least) one Majorana fermion. The sector of the Lagrangian that describes its dynamics is
\be
\label{Lagrangian interacting theory}
\mathcal{L}=i\overline{N}\slashed{\partial}N-\frac{M}{2}(\overline{N^{c}}N+\overline{N}N^{c})-\overline{N}\slashed{B}\gamma^{5}N-Y_{k}\overline{L}_{k}\tilde{\phi}N+h.c.
\ee
$N$ is the Majorana field and $L_{k}$ is a lepton field, with $k$ a generation index. The adjoint of the Higgs field is defined by the relation
\be
\tilde{\phi}_i=\varepsilon_{ij}\phi_j
\ee
Since leptons have definite chirality, the Yukawa interactions can be rewritten as
\be
\mathcal{L}_{YUK}=-Y_{k}\overline{L}_{k}\tilde{\phi}\left(\frac{1+\gamma^{5}}{2}\right)N-Y_{k}^{*}\overline{N}\tilde{\phi}^{\dagger}\left(\frac{1-\gamma^{5}}{2}\right)L_{k}.
\ee
Using the properties of the charge conjugation matrix and the Majorana condition, it is again seen to be equivalent to
\be\label{last Lagrangian}
\mathcal{L}_{YUK}=-Y_{k}\overline{L}_{k}\tilde{\phi}\left(\frac{1+\gamma^{5}}{2}\right)N-Y_{k}^{*}\overline{L}_{k}^{c}\tilde{\phi}^{\dagger}\left(\frac{1-\gamma^{5}}{2}\right)N.
\ee
It is interesting to notice that the two hermitian conjugate terms in the Yukawa Lagrangian are also CPT conjugate. This is to be expected on the basis of the CPT theorem. In fact CPT violation is introduced only by interactions with the background field. Without those, the squared matrix elements obtained from tree level diagrams for the two decays would be the same \cite{Kolb Wolfram}. From the last form of the interaction Lagrangian, Eq. (\ref{last Lagrangian}), it is straightforward to obtain the Feynman rules for the diagrams giving the decay of the Majorana particle in the two distinct channels. It also allows us to use positive frequency spinors both for the incoming Majorana particle and for the outgoing leptons.

 Let us now turn to the study of the tree-level decay processes of a Majorana right-handed neutrino into leptons and Higgs fields, depicted in fig.~\ref{fig:tree}.
The total four-momentum is conserved in the decay. We use $p$ to denote the four momentum of the Majorana particle, $k$ and $q$ for the four-momentum of the Higgs and the outgoing (anti)lepton respectively.
\begin{align}
E_{p,r}&=E_{q,s}+E_{k}\\
\vec{p}&=\vec{q}+\vec{k}
\end{align}
Note that the energy of the fermions displays an explicit dependence on the helicity. Even assuming the decay products to be massless (which is legitimate, since leptons are actually massless in the unbroken electroweak phase and the Higgs mass parameter is expected to be much smaller compared to the other parameters with dimension of mass), kinematics has to be studied case by case, considering all the possible combinations of the external lines helicities. However, the analysis is much easier if one assumes that the right-handed neutrino is initially at rest. A discussion of the general case, along with a method to find approximate solutions, is given in the Appendix \ref{appendix}. In this case  the following relations hold:
\be
E_{p=0}=\sqrt{B_0^2+m^2}, \qquad E_{k}=|\vec{k}|,\qquad 
E_{q,s}=|B_0+\lambda_{s}|\vec{q}||~.
\ee
Momentum conservation also gives $|\vec{k}|=|\vec{q}|$.

We are then lead to consider two distinct cases, depending on the magnitude of the momentum.

\textbf{Case a} $B_0+\lambda_{s}|\vec{q}|>0$
In this case one has that from $s=2$ follows $m^2=0$, hence for the decay of a massive particle only $s=1$ is allowed and
\be
|\vec{q}|=\frac{\Omega-B_0}{2}.
\ee
In the last formula we introduced the quantity $\Omega$, defined as $\Omega=\sqrt{B_0^2+m^2}$.

\textbf{Case b} $B_0+\lambda_{s}|\vec{q}|<0$
In this case the signs are reversed, so that from $s=1$ follows $m^2=0$. Therefore for the decay of a massive particle $s=2$ is the only allowed case and
\be
|\vec{q}|=\frac{\Omega+B_0}{2}.
\ee

We can finally turn to the calculation of the decay amplitudes, starting with the process $N\rightarrow l^{-}\overline{\phi}$. $U^{r}$ will denote the spinor wave-function of the decaying particle and $u^s$ that of the lepton produced by the decay.   
\begin{align}
\mathcal{M}^{rs}=-iY\overline{u}^s(q)\left(\frac{1+\gamma^{5}}{2}\right)U^{r}(p)=-iY\xi_{s}^{'\dagger}\sqrt{q_{s}\cdot\sigma-B_0}\sqrt{p\cdot\overline{\sigma}+B_0}\,\xi_r=\\
-iY\xi_{s}^{'\dagger}\xi_r\sqrt{E_{q,s}-|\vec{q}_{s}|\lambda_s-B_0}\sqrt{E_{p,r}+B_0+\lambda_{r}|\vec{p}|}\label{decay amplitude}.
\end{align}
The notations $q_{s}$, $E_{s}$  are used to stress the dependence on the helicity  of the four-momentum of the outgoing lepton, and similarly for the incoming particle. Helicity eigenstates corresponding to the outgoing lepton are primed. This is necessary since the momenta $\vec{p}$ and $\vec{q}$ are not parallel, which amounts in our formalism to the use of two distinct axes for the quantisation of the two spins. It is useful for what follows to calculate the scalar products of the two spinors appearing in (\ref{decay amplitude}). Choosing the following helicity eigenstates for the decaying particle (spin along the third spatial direction)
\be
\xi_{2}=\left(\begin{array}{c} 0\\1\end{array}\right), \quad \xi_{1}=\left(\begin{array}{c} 1\\0\end{array}\right).
\ee
One has instead, for the outgoing lepton emitted with angles $\theta,\; \phi$
\be
\xi^{'}_{2}=\left(\begin{array}{c} -e^{-i\phi}\sin\theta/2\\ \cos\theta/2\end{array}\right), \quad \xi^{'}_{1}=\left(\begin{array}{c} \cos\theta/2\\e^{i\phi}\sin\theta/2\end{array}\right).
\ee
Since $E_{q,s}= \left\rvert B_0+\lambda_{s}|\vec{q}_{s}|\right\rvert$ we have to consider two cases, in the same way as we did for the kinematics.

\textbf{Case a} ($B_0+\lambda_{s}|\vec{q}|>0$)
In this case the first square root in (\ref{decay amplitude}) vanishes identically, leading to
\be\label{case a, first decay}
\mathcal{M}^{rs}=0.
\ee

\textbf{Case b} ($B_0+\lambda_{s}|\vec{q}|<0$)
\be
\mathcal{M}^{rs}=-iY\xi_{s}^{'\dagger}\xi_r\sqrt{-2(B_0+\lambda_{s}|\vec{q}_s|)}\sqrt{E_{p,r}+B_0+\lambda_{r}|\vec{p}|}.
\ee

In the case in which the right-handed neutrino is at rest, one knows from kinematics that only $s=2$ is allowed and 
\be
|\vec{q}|=\frac{\Omega+B_0}{2}.
\ee
Therefore
\begin{align}
\mathcal{M}^{r2}=-iY\xi_{2}^{'\dagger}\xi_r\sqrt{-2\left(B_0-\frac{\Omega+B_0}{2}\right)}\sqrt{\Omega+B_0}=\\-iY\xi_{2}^{'\dagger}\xi_r\sqrt{(\Omega-B_0)(\Omega+B_0)}=-iYm\xi_{2}^{'\dagger}\xi_r.
\end{align}
It is important to stress that, as one can see from the last formula, when the spatial part of the total momentum vanishes the decay amplitude is just the standard one.

Calculations for the conjugate decay channel $N\rightarrow l^{+}\phi$ are completely analogous to the previous ones.

The transition amplitude is given by
\be
\mathcal{M}^{rs}=-iY^{*}\overline{u}^s(q)\left(\frac{1-\gamma^{5}}{2}\right)U^{r}(p)=-iY^{*}\xi_{s}^\dagger\sqrt{q_{s}\cdot\overline{\sigma}+B_0}\sqrt{p\cdot\sigma-B_0}\,\xi_r.
\ee
It is non-vanishing only in case \textbf{a}, and it reduces to
\be
\mathcal{M}^{rs}=-iY^{*}\xi_{s}^{'\dagger}\xi_r\sqrt{2(B_0+\lambda_{s}|\vec{q}_s|)}\sqrt{E_{p,r}-B_0-\lambda_{r}|\vec{p}|}.
\ee
In the special case when $\vec{p}=0$ (remember that case \textbf{a} implies that only $s=1$ is allowed) this expression simplifies to
\be
\mathcal{M}^{r1}=-iY^* m\xi_{1}^{'\dagger}\xi_r.
\ee
We next proceed to discuss Leptogenesis induced by a constant $B^0$ background (\ref{defB}), induced by H-torsion in string-cosmology~\cite{Ellis-Nanopoulos-etc}.

\section{H-torsion-background-induced Leptogenesis \label{sec:lepto}}

In this  section we proceed in calculating the relevant quantities needed for an estimate of the lepton asymmetry induced by the torsion background 
within the framework of the Lagrangian (\ref{Lagrangian interacting theory}).

\subsection{Estimation of the lepton abundance \label{sec:abund}}

The relevant quantity for cosmological applications of the theory considered is the thermally averaged decay rate \cite{Kolb-Turner}. This is given by
\be\label{TA decay rate}
\sum_{rs}\int \mbox{d}\Pi_{N,r} \mbox{d}\Pi_{l,s}\mbox{d}\Pi_{\phi} f_N(p_N,r)(2\pi)^4\delta^{(4)}(p_{N,r}-p_{l,s}-p_\phi) |M^{rs}(N\rightarrow l\phi)|^2,
\ee
where we used the following notation for the Lorentz-invariant measure
\be
\mbox{d}\Pi_{X,r}=\frac{\mbox{d}^3 p_{X}}{2E_{X,r}(2\pi)^3}.
\ee
The momenta in the integrand depend explicitly on the spin of the incoming and outgoing particles; hence we separately evaluate each term in the sum (weighted by the respective distribution function). Moreover,  transforming quantities to the centre of mass frame, in principle would introduce the spatial components of $B_{\mu}$. Thus evaluating the integrals in the laboratory frame is preferred. Since we will be considering temperatures lower than the mass of the decaying particle it will be a good approximation to consider the decaying particle to be at rest.

The zero temperature decay rate is obtained by integrating the squared amplitude multiplied by a kinematic factor. The latter results from the integration over momenta of the outgoing particles, enforcing energy-momentum conservation through the delta function. This leaves as a result the measure
\begin{align}
\int \mbox{d}\Pi_{l,s}\mbox{d}\Pi_{\phi}(2\pi)^4\delta^{(4)}(p_{N,r}-p_{l,s}-p_\phi)=
\int\frac{\mbox{d}\varpi}{16\pi^2}\frac{|\vec{k}|}{E_q+E_k\left(1+\lambda\frac{B_0}{|q|}\right)\left(1-\frac{|p|}{|k|}\cos\theta\right)}.
\end{align}
Here $\mbox{d}\varpi$ is the solid angle element, $\vec{k}$ is the momentum of the Higgs particle and $\vec{q}$, $\lambda$ are respectively the lepton momentum and helicity. When $\vec{p}=0$ this formula reduces to
\be
\int\frac{\mbox{d}\varpi}{16\pi^2}\frac{|\vec{k}|}{\Omega+\lambda B_0}.
\ee

We now make the simplifying assumption that the decaying particle is at rest, which is a good approximation for temperatures $T$ satisfying $T\leq m$.
From four-momentum conservation follows
\be
|\vec{k}|=|\vec{q}|=\frac{\Omega-\lambda B_0}{2},
\ee
where $\lambda$ is the helicity of the (anti)lepton produced in the decay and $\Omega=\sqrt{B_0^2+m^2}$ is the energy of the initial particle. It is worth noting that only the case $|\vec{q}|+\lambda B_0>0$ is allowed for the decay of a massive particle at rest, since the opposite sign in the inequality implies that $m^2=0$ (the instability of massless particles is a peculiar feature of Lorenz violating theories but is not relevant for our model).
In this special case one has for both channels $N\rightarrow l^{-}\overline{\phi}$ and $N\rightarrow l^{+}\phi$  that the squared matrix element, averaged over the initial spin, has the value $|Y|^2m^2/2$. This seems to lead to a trivial result, implying that it is impossible to generate a lepton asymmetry with this mechanism when the temperature drops to a value comparable to the energy of the decaying particle. However, this conclusion is incorrect, since there is a non-trivial dependence of the kinematic factor on the background field.
We have for the channel $N\rightarrow l^{-}\overline{\phi}$
\be
\Gamma_1=\sum_k\frac{|Y_k|^2}{32\pi^2}\frac{m^2}{\Omega}\frac{\Omega+B_0}{\Omega-B_0},
\ee
while the decay rate for the other channel, $N\rightarrow l^{+}\phi$,  is
\be
\Gamma_2=\sum_k\frac{|Y_k|^2}{32\pi^2}\frac{m^2}{\Omega}\frac{\Omega-B_0}{\Omega+B_0}.
\ee
It is interesting to see that the decay rate of one process is obtained from the other upon flipping the sign of $B_0$.
The total decay rate is
\be
\Gamma=\Gamma_1+\Gamma_2=\sum_k\frac{|Y_k|^2}{16\pi^2}\frac{\Omega^2+B_0^2}{\Omega}.
\ee
It is worthwhile observing that this mechanism can produce a lepton asymmetry even with \emph{only one right-handed neutrino}, whereas the standard leptogenesis scenario \cite{Fukugita-Yanagida} requires at least three generations. Moreover, the occurrence of leptogenesis here is just due to decay processes at tree level, since the required $CP$ violation is introduced by the background field that enters in the external lines of Feynman diagrams.

The decay process goes out of equilibrium when the total decay rate drops below the expansion rate of the Universe, which is given by the Hubble constant \cite{Weinberg}
\be
\Gamma\simeq H=1,66 T^2 \mathcal{N}^{1/2} m_{P}^{-1}.
\ee
Here $\mathcal{N}$ is the effective number of degrees of freedom of all elementary particles and $m_{P}$ is the Planck mass.
From the last equation one can estimate the decoupling temperature $T_{D}$ in terms of the unknown parameters $\Omega$, $|Y|$ and $B_0$.
\be
T_{D}\simeq6.2\cdot 10^{-2} \frac{|Y|}{\mathcal{N}^{1/4}}\sqrt{\frac{m_{P}(\Omega^2+B_0^2)}{\Omega}}
\ee
In order for the inverse decay to be suppressed by the Boltzmann factor, we have to impose the further requirement that $T_{D}\leq \Omega$ when $\Gamma\simeq H$ (delayed decay mechanism \cite{Weinberg, Weinberg 2, Fukugita-Yanagida}). From this condition one can determine a lower bound for the mass $m$. In fact we are lead to the following inequality
\be
z(\Omega^2+B_0^2)\leq\Omega^3,
\ee
where $z=3.8\cdot 10^{-3}\frac{m_{P}|Y|^2}{\mathcal{N}^{1/2}}$. If we require that the bound is satisfied for all values of $B_0$ we get
\be\label{lowerb}
m^2\geq 1.09\, z^2.
\ee
In general, in our scenarios the Yukawa coupling  $Y$ is a free parameter. 
If we assume $|Y|\approx 10^{-5}$, $\mathcal{N}\approx 10^2$, we get an order of magnitude estimate for the lower bound $\overline{m}\approx 100 \;\mbox{TeV}$. 

The lepton number density produced can then be estimated in the following way. We assume that all the neutrinos are at rest before the decay, hence with branching ratios given by $r=\frac{\Gamma_1}{\Gamma}$ and $1-r$. The decay of a single neutrino produces the lepton number
\be \label{lepton number production in one decay}
\Delta L=r-(1-r)=2r-1=\frac{2\Omega B_0}{\Omega^2+B_0^2}.
\ee
Multiplying this quantity by the initial abundance of right-handed Majorana neutrinos at the temperature $T_D$ one gets a rough estimate of the lepton number density. The density of the Majorana neutrinos is given by
\be \label{right handed density}
n_{N}=\sum_{\lambda}\frac{1}{(2\pi)^3}\int\mbox{d}^3 p \, f(p,\lambda).
\ee
where as usual $\beta$ is the inverse temperature, $\lambda$ denotes the helicity and $f(p,\lambda)$ is the corresponding Fermi-Dirac distribution function. At high temperatures this is well approximated by the Maxwell-Boltzmann function. Therefore we set
\be
f(p,\lambda)=e^{-\beta\sqrt{m^2+(p+\lambda B_0)^2}}.
\ee
We can rewrite (\ref{right handed density}) as
\be\label{rewritedensity}
n_{N}=\frac{1}{2\pi^2}\sum_{\lambda}\left(\,I_{2}(-\lambda B_0,\beta,m)-2\lambda B_0\,I_1(-\lambda B_0,\beta,m)+B_0^2\,I_0(-\lambda B_0,\beta,m)\,\right).
\ee
The functions in round braces are defined as follows:
\be
I_{n}(a,\beta,m)=\int_{a}^{\infty}\mbox{d}p\; p^{n} e^{-\beta\sqrt{m^2+p^2}}.
\ee
Retaining only terms that are at most linear in $B_0$ we get that the term proportional to $I_0$ drops and $I_1$ can be evaluated at the zero-th order in $B_0$\footnote{For an alternative derivation that doesn't involve Erf functions, one can consider the approximate expression 
\begin{equation*}
I_{n}\left(a,\beta,m\right)\simeq\int_{0}^{\infty}dp\:\left(p+a\right)^{n}f\left(p\right)\exp\left(-\beta\, p\right)
\end{equation*}
where $f(p)=e^{-\beta\sqrt{a^{2}+m^{2}}}\left[1+p\left(\beta-\frac{\alpha\beta}{\sqrt{\alpha^{2}+m^{2}}}\right)+\frac{1}{2}p^{2}\left(\left(\beta-\frac{\alpha\beta}{\sqrt{\alpha^{2}+m^{2}}}\right)^{2}-\frac{\beta m^{2}}{\left(\alpha^{2}+m^{2}\right)^{3/2}}\right)\right]$.}. Moreover we have:
\begin{align}
I_1(0,\beta,m)&=\frac{1+\beta m}{\beta^2}\,e^{-\beta m}\\
I_2(-\lambda B_0,\beta,m)&=e^{-\beta m}\Big[\frac{-\lambda B_0 \, m}{\beta}+\sqrt{\frac{\pi}{2}}\left(\frac{m}{\beta}\right)^{\frac{3}{2}} \mbox{Erfc}\left(-\lambda B_0\sqrt{\frac{\beta}{2m}}\right)\Big]\label{nonrelintegral}
\end{align}
The last formula, Eq.(\ref{nonrelintegral}), is valid in the non-relativistic limit $\sqrt{m^{2}+p^2}\simeq m+\frac{p^2}{2m}$. The complementary error function is defined as the integral of the Gaussian function
\be
\mbox{Erfc}(z)=\frac{2}{\sqrt{\pi}}\int_{z}^{\infty}\mbox{d}u\; e^{-u^2}.
\ee
Since its derivative is given by
\be
\mbox{Erfc}^{\prime}(z)=-\frac{2\, e^{-z^2}}{\sqrt\pi},
\ee
one has that, expanding around $B_0=0$, (\ref{nonrelintegral}) reduces to
\be
I_2(-\lambda B_0,\beta,m)=e^{-\beta m}\,\sqrt{\frac{\pi}{2}}\left(\frac{m}{\beta}\right)^{\frac{3}{2}}+\mathcal{O}(B_0^2).
\ee
It is now straightforward to see that, performing the sum over helicities in (\ref{rewritedensity}), one recovers the usual expression for the density of a non-relativistic species
\be
n_{N}=e^{-\beta m}\,\left(\frac{m}{2\pi\beta}\right)^{\frac{3}{2}}+\mathcal{O}(B_0^2).
\ee
We assume that the right-handed neutrino density distribution follows closely the equilibrium distribution for $T\geq T_{D}$ and drops rapidly to zero at lower temperatures $T\leq T_{D}$; furthermore  the density of the sterile neutrino (normalised to the entropy density) is well approximated by a step-function. Therefore we have, upon multiplying (\ref{lepton number production in one decay}) by $n_{N}$, that the total lepton asymmetry produced in the full decay of the right-handed neutrino is given by
\be
\Delta L^{TOT}=(2r-1)n_{N}=\frac{2\Omega B_0}{\Omega^2+B_0^2}n_N
\ee
The lepton asymmetry $\frac{\Delta L^{TOT}}{n_{\gamma}}$ is expected to be of the same order of magnitude of the baryon asymmetry (\ref{baryon asymmetry}).
An order of magnitude estimate of the ratio $\frac{B_0}{m}$ can be found making use of the approximation $T_D\simeq m$ and retaining only first order terms in $\frac{B_0}{m}$.

Recalling that the photon number density is
\be
n_\gamma \simeq \frac{2\zeta(3)}{\pi^2}\,T^3\simeq 0.24\, T^3
\ee
and that
\be
\frac{\Delta L}{n_\gamma}\simeq 10^{-10},
\ee
We can estimate the ratio of the background field to the mass of the sterile neutrino to be
\be\label{largeB}
\frac{B_0}{m}\simeq 10^{-8}.
\ee
The small value of this ratio also allows us to justify \emph{a posteriori} neglecting higher powers of $B_0$ in the formulae above. From the lower bound (\ref{lowerb}) of 100 TeV that has been previously found  for the mass, for the case where $Y = {\mathcal O}(10^{-5})$, we get an approximation for the smallest possible magnitude of the background field required in order for this mechanism to be effective $B_0\simeq1 \; \mbox{MeV}$.  
If other mechanisms contributed to the lepton asymmetry in the universe, or the Yukawa couplings assume smaller values, the minimum value of $B_0$ would be smaller than the one given here.

  A $B_0$ of this magnitude would correspond, on account of the effective action (\ref{binteg}) 
to a large positive contribution to the cosmological constant, which would modify the standard cosmology in the radiation dominated eras of the early Universe, that so far we used. However, this is not the case in the context of higher-dimensional string effective models, say formulated on brane universe. In such cases, there are negative (anti de-Sitter) type contributions to the brane (four-dimensional) vacuum energy coming from the extra-dimensional bulk. Such contributions may suppress the $B_0$-induced vacuum energy contributions to acceptable levels so that the standard cosmology may apply (\emph{cf}. Eq.~(\ref{low energy effective action})).

Moreover, a large $B_0$ of order (\ref{largeB}) should be absent today, given that precision atomic experiments have placed stringent upper bounds on $B_0 \le {\mathcal O}(10^{-2})$~eV, within the context of experimental tests of the Standard Model Extension~\cite{Colladay-Kostelecky,bounds}.
In such a case one has to assume that the torsion field was present in the early Universe but underwent a phase transition at a certain stage after the above decoupling temperature $T_D$, so that it is practically absent or very small today. 
One way to think of the nature of the associated transition is to assume that it occurs at a critical temperature $T_c \le T_D $, so that 
for $T < T_c$ the chiral current condensate vanishes, in which case the axion field $b$ no longer varies linearly with the cosmic time but rather diminishes with the scale factor as in (\ref{bscale}). It is then easy to see that, if one assumes a cooling law for the Universe, of the form $a \sim T^{-1}$, then the $B^0$ torsion field would scale with the temperature  as $T^3$, for $T \le T_D$ in this scenario. Taking into account that today, 
the temperature of the Universe (that of CMB) is $T_{\rm CMB} = 2.725~{\rm K} = 0.2348 ~{\rm m eV} $, and assuming that the phase transition occurs at temperatures of the order of Leptogenesis, that is $T \simeq T_D = 100$~TeV, we obtain a cooling law for the torsion $B_0$-field of the form
\begin{equation}
B_0 =  c_0 \, T^3~, \quad c_0 = {\rm 1~MeV}{(100~{\rm TeV})^{-3}} = 10^{-42}\, {\rm meV}^{-2}~.
\end{equation}
Thus, today the value of $B_0$ is such a scenario would be of order 
\begin{equation}
B_{0 \, \rm today} = {\mathcal O}\Big(10^{-44}\Big)~{\rm meV}~, 
\end{equation}
way too small for any experimental detection. 

In this scenario, to keep the cosmological vacuum energy on the brane world small after the phase transition, so that the standard cosmology can apply, we have to assume that any bulk anti-de-Sitter contribution also varies with the cosmic time so as to compensate any effects coming from $b(t)$ in this case. This is plausible in string models,\emph{ e.g}. by assuming that the brane universe, as it moves into the bulk space, encounters different densities of effectively ``point-like'' bulk brane defects; interactions of the brane universe with such defects occur via stretched open strings between the brane Universe and the bulk defects, which are known to be of mixed sign~\cite{dflation}.

However, we should point out that, in order to get a more accurate estimate of $B_0$,  the relevant Boltzmann equation needs to be studied. This requires a good approximation for the thermally averaged decay rates (\ref{TA decay rate}) of all the relevant processes and will be the subject of future research. Nevertheless, in Appendix \ref{sec:appendix2} we construct the Boltzmann equation, with the simple purpose of demonstrating the differences induced by the background $B_0 \ne 0$ and the challenges one faces when attempting to solve it.

\subsection{A comment on equilibrium leptogenesis in this framework}
The coupling of fermions to the axial field $B_{\mu}$ induces, as we have seen in the preceding sections, different dispersion relations for states with opposite helicity. The density of a given particle species is indeed given by
\be
n=\frac{g}{(2\pi)^3}\int\mbox{d}^3 p f(p),
\ee
where $g$ is the number of degrees of freedom and $f(p)$ the probability distribution function in momentum space. For fermions this is a Fermi-Dirac function
\be
f(p)=\frac{1}{\exp{\frac{E(p)}{kT}}+1}.
\ee
If we have that the Lorentz-violating interactions give the leptons and antileptons different energies (in an analogous fashion to what was done in \cite{Colladay-Kostelecky}, where equilibrium baryogenesis is considered instead of leptogenesis) for corresponding values of the momentum and helicity quantum numbers, then the lepton asymmetry can be calculated as
\be
n_{L}-n_{\overline{L}}=\frac{g}{(2\pi)^3}\int\mbox{d}^3 p \; (f_{L}(p)-f_{\overline{L}}(p)).
\ee
In  \cite{Colladay-Kostelecky} interactions are considered that lead to a uniform shift of the energy levels. The shift can be interpreted as a chemical potential, that happens to be different for particles and antiparticles due to CPT violation. However, the  present case is different, since particles and antiparticles with the same helicity also have the same dispersion relation, hence the same density. There is a difference in density just between positive and negative helicity states, regardless of the fact that they belong to the same or to different particle species. For this reason there is no lepton asymmetry at equilibrium that can be justified on the basis of the CPT violating interaction in the Lagrangian (\ref{diracb}).

However, there is an asymmetry at equilibrium between right-handed neutrinos and anti-neutrinos that can be interpreted as a particle-antiparticle asymmetry. The right handed neutrino is a weak isospin singlet, and is therefore allowed to have a Majorana mass. Since a Majorana particle is C-conjugated, the only way to distinguish a neutrino from its antineutrino is via CP conjugation, or equivalently by the helicity. Therefore the asymmetry between opposite helicity right-handed neutrino states amounts to an asymmetry between the density of neutrinos and antineutrinos. However, this cannot be interpreted in terms of a lepton asymmetry, since the right handed Majorana neutrino has no definite lepton number. The way this asymmetry might contribute to leptogenesis is only through the decay of the right-handed neutrino, \emph{i.e.} when this neutrino states are converted to other states having a definite lepton-number. This is indeed the mechanism considered in our paper.

\section{Conclusions and Outlook \label{sec:concl}} 

In this work we elaborated further in the programme we embarked a couple of years ago~\cite{Mavromatos-Sarkar,Environmental} in order to discuss issues associated with the r\^ole of string-inspired Kalb-Ramond torsion on Leptogenesis, and subsequent Baryogenesis, e.g. through sphaleron processes.  Unlike the case of torsionless Riemannian manifolds, the presence of torsion, and especially the one associated with the totally antisymmetric tensor field strength in string theory, may imply for certain string backgrounds, a lepton-number asymmetry in the early Universe, as a consequence of different decay rates of heavy right-handed neutrinos in the early universe into leptons and antileptons. This difference is induced exclusively by the torsion, which in certain exact string backgrounds can be constant in the Robertson-Walker frame and it is essentially a tree-level CP violating asymmetry, in contrast to the standard approach to leptogenesis where in order to obtain such asymmetries one has to go to one loop level.

Our simplified model estimates, which have been done in flat space times, gave only an order of magnitude estimate of the induced asymmetry.
More detailed estimates, associated with solving exactly the pertinent Boltzmann equations in the presence of torsion, are left for a future work.
Nevertheless, for completeness, in the current article we also sketched the modifications induced by the torsion field in the collisionless Boltzmann equation and derived the associated particle distribution function, which was found to be well behaved for non zero values of the temporal component of the axial vector background due to torsion, which we consider here. 

Our simplified model for leptogenesis involves a single flavour of a heavy Majorana neutrino and the associated Yukawa coupling $Y$ 
that couples it to the standard model lepton sector. Although, if one considers seesaw mechanisms for mass generation of the active left-handed neutrinos, one concludes that (s)he needs three right-handed neutrinos as well, nevertheless, the presence of stringy torsion implies alternative scenarios for left-handed neutrino Majorana mass generation, such as the one in \cite{MavPilaftsis}, in which mixing of torsion pseudoscalar field fluctuations with ordinary axion fields that in turn couple, via appropriate Yukawa interactions, to the active left-handed neutrino sector,  can produce dynamical generation of Majorana neutrino masses via higher loop anomalous graphs. In such generalised scenarios, the parameter $Y$, the mass of the heavy right handed neutrino and the value of the constant background antisymmetric tensor field strength  are free parameters to be determined by the requirement of generating the correct lepton asymmetry. Embedding of such scenarios in detailed microscopic string models may 
lead to restrictions on the allowed constant values of the torsion.

From the crude estimates presented in the present work, it is apparent that in order to reconcile the current experimental evidence on the absence of appreciable effects of torsion or the associated Lorentz and CPT violating axial background field $B_0$ in the current era~\cite{bounds}, 
phase transitions of the Kalb-Ramond field at some stage after decoupling should be envisaged. Details of such transitions depend on the string theory microscopic cosmological models in which our scenario for leptogenesis is embedded. 
We do hope to be able to report on some progress in this latter direction in the future. 

\emph{Affaire \`a suivre... }

\appendix

\section{Fermions in Space-time Backgrounds with Torsion }\label{appendix1}

It is well known~\cite{Shapiro} that a theory of fermions in a space time with torsion (Einstein-Cartan theory) results in a four-fermion interaction after integrating out torsion in a path integral.  This is easily understood by the fact that the torsion is a non-propagating field in the Einstein Cartan theory,
where the gravitational field dynamics is described only by a generalised scalar curvature term coupled to Dirac fermions (which may or may not be charged)
\begin{eqnarray}\label{dirac}
S_{\rm EC} &=& \frac{1}{8\pi{\rm  G}} \, \int d^4 x \sqrt{-g}\left( \widebar{R}(\widebar \omega)  -  S_\psi\right)~, \nonumber \\
S_\psi &=& \frac{i}{2} \int d^4 x \sqrt{-g} \Big( \widebar{\psi}
\gamma^\mu \widebar{\mathcal{D}}_\mu \psi  
- (\widebar{\mathcal{D}}_\mu \widebar{\psi} ) \gamma^\mu \psi \Big)
\end{eqnarray}
where $\widebar{\mathcal{D}}_\mu = \widebar{\nabla}_\mu - i e A_\mu $,
is the covariant derivative of 
with  $e$ the  fermion  charge  and $A_\mu$ an electromagnetic  field.   
The overline above  the covariant derivative, i.e.~$\widebar{\nabla}_\mu$,
denotes  the presence  of  torsion, which  is  introduced through  the
torsionful spin connection: 
\be
\widebar{\omega}_{a \mu b } = \omega_{a \mu b } + K_{a   \mu b}~, 
 \ee 
where $K_{a  \mu b}$  is the contorsion tensor (as usual greek letters denote components in the coordinate basis, while latin indices refer to the tetrad).
The  latter  is  related  to  the  torsion  two-form $\textbf{T} =
\textbf{d   e}  +   \boldsymbol{ \widebar{\omega}}  \wedge   \textbf{e}  $
via~\cite{kibble,Shapiro,kaloper}:     
\be
K^{\lambda}_{\;\mu\nu}    =     \frac{1}{2}    \Big(
\textrm{T}^{\lambda}_{\;\mu\nu}  + \textrm{T}_{\mu\;\nu}^{\;\lambda} + \textrm{T}_{\nu\;\mu}^{\;\lambda}  \Big)~.  
\ee
The
presence  of torsion  in the  covariant derivative  in  the Dirac-like
action (\ref{dirac}) leads, apart from the standard terms in manifolds
without  torsion, to an  additional term  involving the total axial current (the sum runs over all fermion species)
$J^\mu_5 \equiv  \sum_{k}\widebar \psi_{k} \, \gamma^\mu \, \gamma^5 \, \psi_{k} $:
\begin{equation}\label{torsionpsi}
S_\psi \ni  - \frac{3}{4} \int d^4 \sqrt{-g} \, S_\mu \widebar{\psi}
\gamma^\mu \gamma^5 \psi  = - \frac{3}{4} \int \mathbf{S} \wedge {}^\star\! \mathbf{J}^5  
\end{equation}
where $\textbf{S} = {}^\star\! \textbf{T}$  is the dual of \textbf{T}: $S_d
=   \frac{1}{3!}     \epsilon^{abc}_{\quad   d}   T_{abc}$.   In (\ref{torsionpsi}), and in what follows, we adopt for notational convenience a form language to describe the effective action of fermions in a curved space-time with torsion.  

We  next remark that  the torsion  tensor can  be decomposed  into its
irreducible parts~\cite{Shapiro},  of which $S_d$  is the pseudoscalar
axial vector:
\begin{equation}
T_{\mu\nu\rho} = \frac{1}{3} \big(T_\nu
g_{\mu\rho} - T_\rho g_{\mu\nu} \big) - \frac{1}{3!}
\epsilon_{\mu\nu\rho\sigma} \, S^\sigma + q_{\mu\nu\rho} ~, 
\end{equation}
with
$\epsilon_{\mu\nu\rho\sigma} q^{\nu\rho\sigma} = q^\nu_{\,\rho\nu} =
0$.
This implies that the contorsion tensor undergoes the following decomposition:
\begin{equation}\label{hatted}
K_{abc} = \frac{1}{2} \epsilon_{abcd} S^d + {\widehat K}_{abc} 
\end{equation}
where $\widehat  K$ includes the  trace vector $T_\mu$ and  the tensor
$q_{\mu\nu\rho}$ parts of the torsion tensor.
 
The gravitational part of the action can then be written as:
\begin{equation}\label{toraction}
S_G =\frac{1}{2\kappa^2} \, \int d^4 x \sqrt{-g} \Big(R +
\widehat{\Delta} \Big) + \frac{3}{4\kappa^2} \int \textbf{S} \wedge
{}^\star\! \textbf{S}\; ,
\end{equation}
where  $\widehat \Delta  = {\widehat  K}^\lambda_{\ \: \mu\nu} {\widehat
  K}^{\nu\mu}_{\quad \lambda}  - {\widehat K}^{\mu\nu}_{\quad  \nu} \,
{\widehat K}^{\quad  \lambda}_{\mu\lambda}$, with the  hatted notation
defined in (\ref{hatted}).

In a  quantum gravity setting,  where one integrates over  all fields,
the torsion terms  appear as non propagating fields  and thus they can
be integrated out exactly. The authors of \cite{kaloper} have observed
though   that  the   classical  equations   of  motion   identify  the
axial-pseudovector torsion field $S_\mu$ with the axial current, since
the torsion equation yields
\begin{equation}\label{torsionec}
K_{\mu a b} = - \frac{1}{4} e^c_\mu \epsilon_{a b c d} \widebar{\psi}
\gamma_5 \gamma^d \psi\ .
\end{equation}
From this  it follows $\textbf{d}\,{}^\star\!\textbf{S}  = 0$, leading
to a  conserved ``torsion charge'' $Q =  \int {}^\star\!  \textbf{S}$.
To  maintain  this conservation  in  quantum  theory, one has to postulate
\be\label{bianchitorsion}
\textbf{d}\,{}^\star\!\textbf{S} = 0~,
\ee
 at the \emph{quantum} level, which can
be  achieved  by  the  addition  of  judicious  counter  terms (see \cite{kaloper}).   This
constraint, in a path-integral formulation of quantum gravity, is then
implemented  via a delta  function constraint,  $\delta (\textbf{d}\,{}^\star\!
\mathbf{S})$, and the latter via the well-known trick of introducing a
Lagrange multiplier  field $\Phi (x)  \equiv (3/2\kappa^2)^{1/2} b(x)$.
Hence, the relevant torsion  part of the quantum-gravity path integral
would include a factor {\small
\begin{eqnarray}
 \label{qtorsion}
{\mathcal Z} &\propto& \int D \textbf{S} \, D b   \, \exp \Big[ i \int
    \frac{3}{4\kappa^2} \textbf{S} \wedge {}^\star\! \textbf{S} -
      \frac{3}{4} \textbf{S} \wedge {}^\star\! \textbf{J}^5  +
      \Big(\frac{3}{2\kappa^2}\Big)^{1/2} \, b \, \textbf{d} {}^\star\! \textbf{S}
      \Big] \nonumber \\
      &= & \int D b  \, \exp\Big[ -i \int \frac{1}{2}
      \textbf{d} b\wedge {}^\star\! \textbf{d} b + \frac{1}{f_b}\textbf{d}b 
\wedge {}^\star\! \textbf{J}^5 + \frac{1}{2f_b^2}
    \textbf{J}^5\wedge{}^\star\! \textbf{J}^5 \Big]~,
\end{eqnarray}
where 
\begin{equation}\label{fbdef}
f_b = (3\kappa^2/8)^{-1/2} = \frac{M_P}{\sqrt{3\pi}}, \quad J^{5 \, \mu} = \sum_{k}\widebar \psi_{k} \, \gamma^\mu \, \gamma^5 \, \psi_{k}~, 
\end{equation}
and  the  non-propagating   $\textbf{S}$  field  has  been  integrated
out. The reader  should notice that, as a  result of this integration,
the   corresponding   \emph{effective}   field   theory   contains   a
\emph{non-renormalizable} repulsive four-fermion axial current-current
interaction.

We  may partially integrate  the second  term in  the exponent  on the
right-hand-side  of (\ref{qtorsion})  and take  into account  the well
known  field theoretic result  that in  QED the  axial current  is not
conserved at the  quantum level, due to anomalies,  but its divergence
is obtained by the one-loop result:
\begin{eqnarray}
   \label{anom}
\nabla_\mu J^{5 \, \mu} \!&=&\!  \frac{e^2}{8\pi^2} {F}^{\mu\nu}
  \widetilde{F}_{\mu\nu}  
- \frac{1}{192\pi^2} {R}^{\mu\nu\rho\sigma} \widetilde
{R}_{\mu\nu\rho\sigma} 
\equiv G(A, \omega)\; .
\end{eqnarray}
Observe that in (\ref{anom}) the torsion-free spin connection has been
used.  This can be achieved by the addition of proper counter terms in
the  action~\cite{kaloper}, which  can  convert the  anomaly from  the
initial    $G(\textbf{A},   \widebar   \omega)$    to   $G(\textbf{A},
\omega)$. Using  (\ref{anom}) in (\ref{qtorsion}) one  can then obtain
for the effective torsion action in QED
\begin{equation}\label{brr}
\int D b\ \exp\Big[ - i \int \frac{1}{2}
    \textbf{d} b\wedge {}^\star\! \textbf{d} b  - \frac{1}{f_b} b
    G(A, \omega)  
+ \frac{1}{2f_b^2} \textbf{J}^5 \wedge ^\star\!\textbf{J}^5 \Big]\; .
\end{equation}
Thus, we observe that  the  torsion lead  to  repulsive four-fermion  interactions
involving  the axial  current. Crucial  to the  above  derivation was,
however, the postulation of the  conservation of the torsion charge at
the  quantum  level,  as   expressed  by  the  constraint  $\textbf{d}
{}^\star\!  \textbf{S} = 0$.  The resulting axion field has originated
from the  Lagrange multiplier field implementing  this constraint.  In
the subsequent  section we consider the cosmological implications of this result. The reader should notice that the form of the action (\ref{brr}) is identical (up to irrelevant proportionality constants in the mixed term of the axion $b$-field to the anomaly $G(A, \omega)$) to the one obtained from string theory considerations
with the Kalb-Ramond field as torsion (\ref{binteg}), thereby establishing the equivalence of the two approaches. 

\section{Kinematics of the decay to first order in the external field}\label{appendix}
In this appendix we illustrate how the kinematics of the decay processes considered in the present paper can be studied using simple perturbative techniques. Let's consider the case $N\longrightarrow l^- \overline{\phi}$ for definiteness. The other case is perfectly analogous. All we need to know is that the total energy and the total momentum are both conserved in the process
\begin{align}
E_{N,r}&=E_{l^-,s}+E_{\overline{\phi}},\\
\mathbf{p}&=\mathbf{q}+\mathbf{k}.
\end{align}
Lorentz violation introduces corrections in the first equation through the dependece of the energies of the decaying neutrino and of the lepton on the helicity and on the background field.
\begin{align}\label{sistema}
E_{N,r}&=\sqrt{m^2 +(p+\lambda_{r} B_0)^2}\\  
E_{l^-,s}&=|B_0+\lambda_s q|=-(B_0+\lambda_s q)
\end{align}
Here $p$ and $q$ represent respectively the norms of the spatial vectors $\mathbf{p}$ and $\mathbf{q}$. 
The last step in the last equation follows from what we said in (\ref{case a, first decay}), where we showed that only if the argument of the absolute value is negative can the amplitude be different from zero. In principle, one can solve this system of equations for $q$, $k$ and $\theta_{pq}$ (the angle formed by the vectors $\mathbf{p}$ and $\mathbf{q}$) in terms of $p$, of the helicities, the external field $B_0$ and the angle $\theta_{pk}$ (formed by $\mathbf{p}$ and $\mathbf{k}$). However, as it stands, it is hard to find a solution for the system of equations (\ref{sistema}).

It is convenient to treat the corrections coming from Lorentz violation as perturbations (which is justified if $B_0\ll m$, $T$) and define the solutions of the system as the sum of the unperturbed ones plus perturbations 
\begin{align}
k&=k^0+k^1,\\
q&=q^0+q^1,\\
\theta_{pq}&=\theta_{pq}^0+\theta_{pq}^1.
\end{align}
Introducing the adimensional parameter $\varepsilon=\frac{B_0}{m}$, that will play the role of an expansion parameter, we write down the energy of the Majorana neutrino to first order in $\varepsilon$
\be
E_{N,r}=E_{N}^0\left(1+\varepsilon\lambda_{r} \alpha \frac{p}{m}\right)+\mathcal{O}(\varepsilon^2).
\ee
Having defined $\alpha=\frac{m^{2}}{(E_{N}^0)^2}$.
We get the linearized system of equations
\begin{align}
k^1-\lambda_{s} \;q^1&= C\\
\sin\theta_{pk}\; k^1 +\sin\theta_{pq}^0 \;q^1+q^0\cos\theta_{pq}^0 \;\theta_{pq}^1&=0,\\
\cos\theta_{pk} \;k^1 +\cos\theta_{pq}^0 \;q^1 - q^0 \sin\theta_{pq}^0 \;\theta_{pq}^1&=0,
\end{align}
which has the following solution
\begin{align}
k^1 &=\frac{C}{1+\lambda_s\cos\left(\theta_{pk}-\theta_{pq}^0\right)},\\
q^1 &=-\frac{C \cos\left(\theta_{pk}-\theta_{pq}^0\right)}{1+\lambda_s \cos\left(\theta_{pk}-\theta_{pq}^0\right)},\\
\theta_{pq}^1 &=-\frac{C\sin\left(\theta_{pk}-\theta_{pq}^0\right)}{q^0\left(1+\lambda_s\cos\left(\theta_{pk}-\theta_{pq}^0\right)\right)}.
\end{align}
The definition of $C$ is the following
\be
C=\varepsilon\lambda_{r} \alpha \frac{p}{m} E_{N}^0+(1+\lambda_s)q^0+B_0.
\ee
In the COM frame $\theta_{pk}=\theta_{pq}^0 + \pi$, implying that only $\lambda=-1$ is allowed and the usual corrispondence between helicity and chirality is re-established. This is in agreement with equation (\ref{case a, first decay}), that was derived independently.

\section{Towards a more quantitative estimate of lepton abundance: Boltzmann Equation in the presence of $B_0 \ne 0$ axial background\label{sec:appendix2}}

In this Appendix we discuss how the qualitative estimates for leptogenesis in section \ref{sec:abund} can be confirmed by following a more accurate approach, specifically that of the Boltzman equation for the thermal relic abundance at decoupling of the right-handed neutrinos of various helicities.
The Boltzmann equation \cite{Kolb-Turner} essentially expresses the action of the so-called Liouville operator $\widehat{L}[f]$ on the phase-space density of the species $\chi$,  $f({\vec x}, |{\vec p}|, t )$,  in terms of the so-called collision operator $C[f]$, monitoring the deviation from equilibrium in the reactions that the species $\chi$ participates (for an application of the Boltzmann equation to simple model of baryogenesis see \emph{e.g.} \cite{Kolb Wolfram 2}). We assume for concreteness one single dominant species, with mass $m_\chi$. 

In the non-relativistic (Newtonian) case, the Liouville operator is a total time derivative, time is universal, and ${\vec x} (t)$, ${\vec p}(t)$ depend on time (phase-space trajectory of the particle):
so its action on
 $f({\vec x}, |{\vec p}|, t )$ is given by:
 \begin{gather}\label{liouvnewt}
\widehat{L}[f] = \frac{d}{dt} f = \frac{\partial}{\partial t} f + {\vec v}\cdot{\vec \nabla} f + \frac{\vec F}{m_\chi}\cdot \nabla_{\vec v} f
\end{gather}
where $\vec v = d\vec x/dt  $ is the velocity, and $\vec F = d\vec p/d t$ is the (Newtomnian) force acting on the particle.

The extension of (\ref{liouvnewt}) to the general-relativistic case, that will allow treatment in the Robertson-Walker Universe, is straightforward. Essentially, the Newtonian total time derivative of the non-relativistic case is replaced by a total \emph{derivative} with respect to the \emph{proper time}.
The resulting Liouville operator is essentially,
\begin{gather}\label{relliouv}
\widehat{L}[f]~ \rightarrow ~  m_\chi \frac{d}{d\tau} f~ = m_\chi  u^\alpha \partial_\alpha f + m_\chi  \frac{d p^\alpha}{d \tau} \frac{\partial}{\partial p^\alpha} f~,
\end{gather}
where $u^\mu$ is the four-velocity and $p^\mu = m_\chi u^\mu$ the four-momentum. In (\ref{relliouv}) we took into account that any dependence of the phase-space density $f$ on the proper time $\tau$ is through the dependence of $x^\mu (\tau), p^\alpha (\tau)$ on $\tau$.

Based on our discussion so far, then, the combination
$\frac{\partial}{\partial t} f + {\vec v}\cdot{\vec \nabla}$  of the Newtonian case is replaced in General Relativity by $u^\alpha \partial_\alpha $, whilst the `force' term is expressed in terms of the Christoffel symbols by means of the geodesic equation, 
\begin{equation}\label{geodesic}
m_\chi  \frac{d p^\alpha}{d \tau} = - \Gamma ^\mu_{\alpha\beta} p^\alpha p^\beta ~.
\end{equation}
Notice that the torsion (antisymmetric part of the Christoffel symbok) does not enter the geodesic equation. Nevertheless, as we shall discuss below,
the equation is still modified by the presence of the $B_0$ vector, due to the modified dispersion relations (\ref{dispersion relation}) for the various helicity states.
 
The result for the general-relativistic Liouville operator is, therefore:
\begin{gather}
\widehat{L}[f] = [p^\alpha \partial_\alpha - \Gamma^\alpha_{\mu\nu}p^\mu p^\nu \frac{\partial}{\partial p^\alpha} ]f~.
\label{grliouv}
\end{gather}
For a homogeneous and isotropic Robertson-Walker Universe, with a scale factor $a(t)$, we have that $f=f(t,|\vec p|) $. Equivalently,
upon using the RW-space-time on-shell condition for the massive species $\chi$, $f=f(E,t)$, where $E$ denotes the energy of the particle and $t$ is the co-moving frame RW cosmic time. On using the Christoffel symbols for the Robertson-Walker metric, 
we obtain from (\ref{grliouv}):
\begin{gather}
\widehat{L}[f] = E \frac{\partial f}{\partial t} - \frac{\dot a}{a}|\vec p|^2 \frac{\partial f}{\partial E}~.
\label{liouvrw}
\end{gather}
The number density of species $n_\chi $ is defined as:
\begin{gather}\label{numberdens}
n_\chi = \frac{g}{8\pi^3}\int d^3p f(E,t)
\end{gather}
where $g$ is the number of degrees of freedom of the species $\chi$. Dividing  (\ref{liouvrw}) by $E$, integrating over all momenta and using the definition (\ref{numberdens}), we obtain:
 \begin{eqnarray}
 \frac{d n_\chi }{d t} - \frac{\dot a}{a} \frac{g}{8\pi^3}\int_{0}^{\infty}  d|\vec p| d\Omega~ \frac{|\vec p|^4}{E}
 \frac{\partial |\vec p|}{\partial E}
 \frac{\partial f}{\partial |\vec p|} 
\label{inter}\end{eqnarray}
where in the last step we have spilt the momentum integration into momentum-amplitude ($|\vec p|$) and angular ($\Omega$) parts, and transformed the $E$-differentiation to a $|\vec p|$-differentiation. Consider now the abundance of a particluar helicity state $\chi = r$, $n_r$. Using the dispersion relations 
(\ref{dispersion relation}) we obtain 
$$ \frac{\partial |\vec p|}{\partial E_r} = \frac{E_r}{|\vec p| + \lambda_r \, B_0}~, \quad \lambda_r = \pm 1 ~.$$
with the notation $|\vec p|^2 \equiv p_i p_j h^{ij}$, where $h_{ij}$ is the spatial part of the RW metric in the standard notation. For concretess  and consistency with astrophysical observations (if one neglects the small value of the cosmological constant), we may assume  that the Universe is spatially flat, in which case $h_{ij}=\delta_{ij}$.

Notice that, depending on the sign of $B_0$, the quantity $|\vec p | + \lambda_r B_0$ may vanish. However, the integrand of (\ref{inter}) is regular, as
$\partial f /\partial |\vec p| \propto |\vec p| + \lambda_r B_0$, for the Boltzmann (thermal) distribution at temperature $T$ $$f(E_r; T) = \frac{1}{e^{E_r/T} + 1}~,$$ assuming zero chemical potential for the relativistic right-handed neutrinos of helicity $r$  for simplicity, with $E_r$ given by (\ref{dispersion relation}). With this in mind, we can expand (\ref{inter}) in powers of (the small compated to the temperature $T_D$) $B_0$ and integrate by parts to arrive at 
a modified Boltzmann equation for the number density of a helicity state $r$ in the form:
\begin{eqnarray}\label{modbol}
&& \frac{g}{8\pi^3}\int \frac{d^3p}{E}C[f] = {\rm d}n_r/{\rm d}t  + \frac{g}{8\pi^3} \, \frac{{\dot a}}{a}\, \int  \, d|\vec p| d\Omega~  \frac{\partial}{\partial |\vec p|} \Big(\frac{|\vec p|^4}{|\vec p| + \lambda_r \, B_0}\Big)\, f  \simeq  \nonumber \\
&& {\rm d}n_r/{\rm d}t + 3Hn_r - \frac{g}{2\pi^2} \, 2\lambda_r \, B_0 \int d|\vec p| \, |\vec p| \, f + {\mathcal O}(B_0^2) ~,\end{eqnarray}
where $ H = \frac{\dot a}{a} $ is the Hubble parameter of the universe.  If we restrict ourselves to small $B_0/T \ll 1$, which is to be expected from our
qualitative estimates in the previous subsection, then we may ingore any $B_0$ dependence of $f$ in the last integral of the right-hand-side of (\ref{modbol}), and thus  replace $f$ by the  
standard Boltzmann distribution of a particle of mass $m$ with energy $E(B_0) = \sqrt{m^2 + |\vec p|^2} $. Passing 
onto dimensionless variable $|\vec p|/T \equiv u$, we then obtain the modified Boltzmann
\begin{eqnarray}\label{modbolfinal}
&&{\rm d}n_r/{\rm d}t + 3Hn_r - \frac{g}{2\pi^2} \, 2\lambda_r \, \frac{B_0}{T}\, T^3  \int du \, u \, f (E(B_0=0), u) 
= \nonumber \\ && \frac{g}{8\pi^3}\int \frac{d^3p}{E(B_0 \ne 0)}C[f] + {\mathcal O}(B_0^2)
\end{eqnarray}
where on the right-hand side one should use the $B_0$ dependent energy.

Eq. (\ref{modbolfinal}) holds for any given species, with a suitable collision operator including contributions for all the interactions it takes part to. In particular, for the right-handed neutrino, we may sum the contributions coming from opposite helicities and get an equation for the total density.
\be\label{boleq}
\frac{\mbox{d} n_{N}}{\mbox{d}t}+3Hn_{N}=\frac{g}{8\pi^3}\int \frac{d^3p}{E}\tilde{C}[f] + {\mathcal O}(B_0^2)
\ee
The collision operator must include contributions for the direct and inverse decays, as well as for the processes with $\Delta L= 1, 2$ considered in \cite{Luty, Buchmuller-Yanagida}. A solution of the appropriate system of Boltzmann equations encountered in our model, including (\ref{boleq}), which arises when the evolution of the abundances of the remaining standard model fields, that also couple to torsion, are taken into account, will be the subject of a forthcoming publication. 

\section*{Acknowledgements}

MdC would like to thank S.~Esposito for correspondence and N.~Houston for useful comments and many fruitful discussions.The work of MdC is supported by King's College London through a GTA studentship. That of NEM is supported in part by the London Centre for Terauniverse Studies (LCTS), using funding from the European Research Council via the Advanced Investigator Grant 267352, and by STFC (UK) under the research grants ST/J002798/1 and ST/L000326/1.

\end{document}